\documentclass{emulateapj}
\usepackage{graphicx}
\usepackage{subfigure}
\usepackage{natbib}
\usepackage{ifpdf}

\shorttitle{Chemistry of WLM}
\shortauthors{Leaman et al.}

\begin{document}

\title{The Comparative Chemical Evolution of an Isolated Dwarf Galaxy: A VLT and Keck Spectroscopic Survey of WLM}
\author{Ryan Leaman$^{1,2,3}$, Kim A. Venn$^{1}$, Alyson M. Brooks$^{4,5}$, Giuseppina Battaglia$^{6}$, Andrew A. Cole$^{7}$, Rodrigo A. Ibata$^{8}$, Mike J. Irwin$^{9}$, Alan W. McConnachie$^{10}$, J. Trevor Mendel$^{1,11}$, Else Starkenburg$^{1,12,13,14}$, Eline Tolstoy$^{12}$}
\affil{$^{1}$University of Victoria, Canada, $^{2}$Instituto de Astrof\'{i}sica de Canarias, Spain, $^{3}$Dept. Astrof\'{i}sica, Universidad de La Laguna, Spain $^{4}$California Institute of Technology, US, $^{5}$University of Wisconsin-Madison, US $^{6}$INAF - Osservatorio Astronomico di Bologna, Italy, $^{7}$University of Tasmania, Australia, $^{8}$Observatoire de Strasbourg, France, $^{9}$Institute of Astronomy, University of Cambridge, UK $^{10}$National Research Council of Canada, Herzberg Institute of Astrophysics, Canada, $^{11}$Max-Planck-Institut f\"{u}r extraterrestrische Physik, Germany, $^{12}$Kapteyn Astronomical Institute, University of Groningen, The Netherlands, $^{13}$CIfAR Junior Fellow, $^{14}$CITA National Fellow}
\email{rleaman@iac.es}

\begin{abstract}
Building on our previous spectroscopic and photometric analysis of the isolated Local Group dwarf irregular (dIrr) galaxy WLM, we present a comparison of the metallicities of its RGB stars with respect to the well studied Local Group dwarf spheroidal galaxies (dSphs) and Magellanic Clouds.  We calculate a mean metallicity of [Fe/H]$ = -1.28 \pm 0.02$, and intrinsic spread in metallicity of $\sigma = 0.38 \pm 0.04$ dex, similar to the mean and spread observed in the massive dSph Fornax and the Small Magellanic Cloud.  Thus, despite its isolated environment the global metallicity still follows expectations for WLM's mass and its global chemical evolution is similar to other nearby luminous dwarf galaxies (gas-rich or gas-poor).  The data also show a radial gradient in [Fe/H] of $\rm{d[Fe/H]/dr_{c}} = -0.04 \pm 0.04$ dex $\rm{r_{c}^{-1}}$, which is flatter than that seen in the unbiased and spatially extended surveys of dSphs.  Comparison of the spatial distribution of [Fe/H] in WLM, the Magellanic Clouds, and a sample of Local Group dSphs, shows an apparent dichotomy in the sense that the dIrrs have statistically flatter radial [Fe/H] gradients than the low angular momentum dSphs.  The correlation between angular momentum and radial metallicity gradient is further supported when considering the Local Group dEs.  This chemodynamic relationship offers a new and useful constraint for environment driven dwarf galaxy evolution models in the Local Group.
\end{abstract}


\section{Introduction}
Dwarf galaxies in the Local Group offer a strong test for mass assembly theories in $\Lambda$CDM cosmologies, as hierarchical merging of protogalactic fragments of similar stellar mass ($10^{5}-10^{9}$ M$_{\odot}$) are expected to be one channel for formation of larger disk galaxies like our own Milky Way (MW) or Andromeda (M31) (e.g., \citealt{NFW97, Moore99, Madau01}).  Characterizing the physical properties of the building blocks of larger galaxies requires an understanding of the global properties of dwarf galaxies such as mass, size, angular momentum, chemistry and luminosity, to observe if the merging fragments are consistent with the properties of the disk and halo of the large spirals (however it is not certain how similar the surviving dwarfs are to those merging fragments; \citealt{Font06}).  

Studies of dwarf galaxies can also shed light on how baryons populate dark matter halos at the faint end of the galaxy luminosity function, and in turn offer a chance to study the star formation and galaxy assembly at low metallicities and masses\citep{BJ05}.  If low star formation efficiency in dwarf galaxies is due to internal feedback effects \citep{Brooks07}, or H2 regulated star formation (i.e., \citealt{KrumDekel11}) will the age and metallicity of a galaxy's stellar population change substantially?  Understanding this question can also provide insight towards what produces the range of morphologies (dwarf spheroidals, transition dwarfs, dwarf irregulars) seen in the Local Group.  Quantitative comparisons of the chemical and kinematic signatures of dwarf galaxies can also constrain evolutionary connections (e.g., tidal transformation scenarios; \citealt{Mayer06}) between the two morphological classes of dwarf galaxies.

Detailed signatures of environmental or internal evolutionary mechanisms can be traced by the evolved stellar populations (RGB stars) in dwarf galaxies.  With spectroscopic observations of these long lived ($\gtrsim 1$ Gyr) stars, questions on whether star formation and chemical enrichment proceed with the same efficiency in isolated and tidally perturbed galaxies can be addressed through analysis of the stellar age-metallicity relation and star formation history (SFH). While the present day appearance and morphology-density relations exhibited by dwarf galaxies in the Local Group and other groups \citep{Weisz11b} indicate that they are susceptible to environmental processing, the details of how environment has influenced them over a Hubble time are still difficult to ascertain.  For example, in a sample of distant dwarf galaxies the recent study of \cite{Weisz11} showed that the SFH of dIrrs and dSphs are nearly identical over the first 12 Gyr and only differ markedly in the most recent 1 Gyr.  However deeper photometric views of lower luminosity Local Group dwarf galaxies find subtle differences between the morphological classes \citep{Hidalgo11}.  The contrasting radial star formation histories between dIrrs and dSphs in that study may be due to differences in internal processes or environment dependent feedback, however disentangling the two and understanding why dIrrs retain more gas to power current SF is difficult.  


The presence of radial abundance gradients and their ubiquity within different classes of dwarf galaxies is one of the primary testing points to infer whether internal or external processes are responsible for the current abundance properties in dwarf galaxies.  Depending on the mechanism for creating, sustaining, and erasing such chemical gradients, their presence and strength may correlate with physical properties or environment of the host system (cf., \citealt{Koleva11}).  At present it is unclear if radial metallicity gradients are ubiquitous in low mass dwarf galaxies of all types, as seen in some dSphs \citep{Saviane01,Harbeck01,Tolstoy04,Battaglia06,Kirby10,Battaglia11}. Therefore it is also unclear if angular momentum and/or radial migration mechanisms mediate these gradients with different efficiencies in galaxies with different dynamical histories, such as the dIrrs \citep{Schroyen11,Roskar08}.

There have been several large scale spectroscopic surveys of the RGB stars in the nearby ($\leq 250$ kpc) dSphs of the Local Group (e.g., \citealt{Battaglia06,Walker09,Kirby10}).  The gas rich dIrrs, lie at distances $500-1100$ kpc away from the MW, which renders analogous surveys observationally expensive.  Studying these isolated dIrrs is informative as they 1) provide important initial structural and kinematic conditions for tidal transformation scenarios which evolve dIrrs into dSphs (cf., \citealt{Kazantzidis11}), and 2) being isolated offer a unique oppourtunity to study internal secular evolution in low mass halos that haven't been strongly perturbed by the tidal forces of the Milky Way.

In the earlier papers in this series (\citealt{Leaman09}; Paper I) and (\citealt{Leaman12}; Paper II) we presented some of the first kinematic analyses of the stellar populations in an extremely isolated dIrr, WLM.  This dwarf galaxy sits 1 Mpc from both the MW and M31, and 250 kpc from the nearest neighbour, the low mass dSph, Cetus.  From its Local Group position and velocity it is inferred that WLM's last pericentre passage was 11-17 Gyr ago, which means it has had at most one close interaction with a massive galaxy \citep{Leaman12}.  The low mass and metallicity of WLM also make it an excellent laboratory for probing chemical evolution in regimes where the SF efficiency is expected to be low \citep{KrumDekel11,Kuhlen11}.  WLM clearly has spent the majority of its lifetime in isolation, and therefore is ideal for disentangling environmental and internal evolutionary processes.  


In this work we present the spectroscopic analysis of the Calcium II triplet (CaT) based [Fe/H] measurements in a sample of 126 red giant branch (RGB) stars in the dIrr galaxy WLM.  As this represents the first isolated Local Group dIrr with a sizable spectroscopic survey of its evolved stellar populations, we proceed with comparing the dynamical, chemical, and structural properties to the less isolated dSphs and Magellanic Clouds.  These comparisons are used to examine the interplay between environmental and internal feedback in an observational sense, and by selecting datasets that have substantially larger spatial extents and unbiased metallicity estimators than past studies, offer a significant improvement in the ability to accurately differentiate the chemical properties of Local Group dwarf galaxies.

\section{Observations and Data Sources}
Resolved stellar spectroscopic data for this paper are from the following sources: spectra of 180 RGB stars were observed with Focal Reducer and low dispersion Spectrograph (FORS2, \citealt{FORS2}) at the VLT (Paper I) and the Deep Imaging Multi-Object Spectrograph (DEIMOS, \citealt{Faber03}) at Keck II (Paper II) for WLM.  Metallicities from FORS2 data were published in paper I, and velocities from both samples were published in paper II. In this paper we determine metallicities from DEIMOS data, as well as re-calibrate the FORS2 EW obtained in paper I.  The dSph RGB spectroscopic data come from the following sources: the Sculptor, Fornax, Sextans, and Carina data are taken from the original surveys of \cite{Tolstoy04, Battaglia06, Battaglia11, Koch06} respectively.  These samples have since been updated by the Dwarf Abundances and Radial velocity Team (DART).  The updates include: observations of additional stars with the same instrument setup and reduction as described in the original papers\footnote{The exception is Carina, whose stars from \cite{Koch06} were taken from the ESO archive and reprocessed using the DART pipeline procedures.}, applying the new CaT-[Fe/H] calibration from \cite{Starkenburg10} to all stars, visual inspection of low metallicity candidates, and quality control cuts (S/N$ > 10$, $\delta_{Vhel} < 5$ km s$^{-1}$).  Throughout this paper spectra from these four galaxies will be referred to as the ``DART sample'', meaning these updated catalogues which are based on the original papers.  The full sample of eight dSphs from \cite{Kirby10} (hereafter K10) were also considered, however as the goal was to characterize the chemical properties of the complete spatial extent in these dwarf galaxies, only Leo I and Leo II are used from that work, as they both show spatial coverage out to at least three core radii (and past the tidal radii). 

The WLM and dSph data are supplemented with observations of RGB \emph{field} stars in the Magellanic Clouds.  For the LMC we draw from the work of \cite{Cole05,Pompeia08,Carrera08b} and the studies of \cite{Carrera08,Parisi10} for the SMC.  The data taken from literature represent spectroscopic surveys of at least 100 stars per galaxy where [Fe/H] estimates are available for each star, and the sample extends to at least $75\%$ of the tidal radius, $r_{t}$, of the galaxy.  In addition WLM and most of the DGs compiled in this work are sampled well along both the minor and major axes, providing a high degree of angular uniformity.  The data sources and references for the nine Local Group dwarf galaxies used in this paper are summarized in Table 1.  We refer the reader to the individual studies for more detailed information on the reductions, but throughout this paper differences in methodology will be discussed when relevant to comparisons we draw. 

\subsection{Auxiliary Measurements}
Additional parameters aside from the published, directly measured velocities and metallicities are required for our analysis.  These are core radii of the dwarf galaxy ($r_{c}$), elliptical radii and ages of the stars.  With these it is possible to explore the spatial and temporal variations that may illuminate differences or connections between the evolutionary history of dIrrs and dSphs.  For the elliptical radii measurements, a galaxy centre, mean ellipticity ($e$) and position angle (PA) are required.  In the case of WLM  the field center from \cite{Mateo98}, and the ellipticity, position angle, and $r_{c}$ from Paper II were adopted.  For Leo I and Leo II dSphs we adopt the field centre from K10 and position angle, $r_{c}$ and $e$ from \cite{IH95}.
Values for the field centre for the DART dSphs were taken from the original papers listed in Table 1 and references therein, and core radii, $e$, and PA from \cite{IH95}.  The field centres, $e$ and PA for the SMC were taken from \cite{HZ06}, and from \cite{vdM01} for the LMC.  The $r_{c}$ for the LMC and SMC were computed as a weighted mean of the values in the recent study of \cite{Belcheva11}.  The geometrical radii for all samples have been derived, and in each case these have been referred to the global ellipticities and position angles listed above.

\section{WLM Spectral Analysis}
For the spectral analysis of WLM we have used the previously observed FORS2 spectra as well as the newly acquired DEIMOS data.  In merging the two datasets possible systematic offsets due to instrumental signatures, equivalent width (EW) measuring techniques and metallicity calibration must be explored.  The following section outlines the influence of the metallicity calibrations on the full sample, with special emphasis on joining the old and new datasets in a way that is consistent within the errors.  This allows for a homogeneous joint sample to be created which has global metallicity properties that are insensitive to instrumental and calibration details.

As in Paper I, [Fe/H] values were derived based on the empirically calibrated Calcium II triplet (CaT) method (e.g., \citealt{ADC91}; \citealt{Rutledge97}).  Originally calibrated with old galactic globular clusters, work has also been done extending the CaT calibration to younger ages \citep{Cole04}, which is relevant here given that the RGB populations in WLM are expected to span 10 Gyrs \citep{Dolphin00}.  However due to the fact that the empirical calibrators (Galactic globular clusters) are only found as metal poor as [Fe/H] $\sim -2.3$, the method has intrinsic limitations.  In recent years work by \cite{Kirby08} has shown the advantage of using synthetic spectral techniques that can derive [Fe/H] in a way that is not limited by the properties of the calibrators - thus pushing down to lower metallicities.  Additionally, the empirical linear calibration of the CaT method has been revised by \cite{Starkenburg10} in order to address the CaT calibration limitation.  They found that at low equivalent widths and magnitudes, the CaT behaviour becomes non-linear, resulting in an overestimate of the [Fe/H] for low CaT equivalent widths.

\subsection{Equivalent Width Measurements of DEIMOS Spectra}
Equivalent width measurements of the Calcium II triplet lines in the new DEIMOS spectra were done with pixel-pixel integration methods, as in Paper I.  For comparison, equivalent width measurements were also produced using integration over the line and continuum bandpasses of \cite{Cenarro01}, and fits to the lines using Gaussian, Lorenztian, Moffat and Voigt functions.  
A comparison of the pixel integration versus several fitted equivalent width estimates for stars in the DEIMOS calibration clusters shows excellent consistency, however the integration bandpasses of \cite{Cenarro01} tends to produce larger values for the EW relative to other methods.  For comparison, the difference in EWs measured by pixel integration and Gaussian plus Lorentzian fits in the lower resolution spectra from Paper I, translated into a 0.17 dex difference in [Fe/H] - which as we will demonstrate is much less than our random uncertainties.  For these data which have relatively low spectral resolutions the direct pixel integration shows the best agreement with Gaussian fits for the calibration stars over all metallicities. However as in Paper I, the resolution and signal-to-noise of the WLM stars necessitate using the integrated equivalent widths rather than functional fits, as the line FWHMs are on the order of the spectral resolution and contaminating noise features nullify any difference between the functional fits.

\subsection{Placement of the Joint Sample onto the Metallicity Scale}
We derive metallicities for WLM using five different CaT-[Fe/H] calibrations to explore the variation in global metallicity properties, these include: \cite{Cole04}, \cite{Battaglia08}, two from \cite{Starkenburg10}, and one based on our DEIMOS calibrating clusters.  Any of these CaT-[Fe/H] calibrations requires a summed equivalent width determination for each star, and are most precise when the EW measurements are done using the same line measuring technique as the original calibrations.  In \cite{Cole04}, each of the three calcium triplet line measurements were combined in an unweighted fashion to yield a summed equivalent width per star $\Sigma W = W_{8498}+W_{8542}+W_{8662}$.  With this relation the calcium index $W' = \Sigma W + \beta(V - V_{HB})$ is formed.  The term in the parentheses provides a correction for 
the changes in T$_{eff}$ and $log(g)$ for stars in different phases on 
the red giant branch.  Our V magnitudes for WLM are taken from the 
INT WFC catalogue presented in \cite{Alan05} and the calibrations in that paper.  
We adopted the horizontal branch at V$_{HB}=25.71\pm0.09$ mag 
\citep{Rejkuba00}, and take $\beta=0.73\pm0.04 {\rm \AA}$ mag$^{-1}$ 
from \cite{Cole04}.  The photometric studies of \cite{Alan05} and \cite{Rejkuba00} find distance moduli of $(m-M)_{0} = 24.85 \pm 0.08$ and $(m-M)_{0} = 24.95 \pm 0.13$ respectively, indicating that there are no significant zero-point offsets that could influence our CaT-[Fe/H] calculations.  Using the \cite{CG97} scale, the calcium index 
is converted to a metallicity ([Fe/H]$_{CG97}$) using Equation 3 of Paper I.

The summed equivalent widths were also converted to an [Fe/H] scale using the calibration from \cite{Battaglia08}, as well as two non-linear calibrations presented in \citep{Starkenburg10}.  In these three cases the summed equivalent width was computed using the two longer wavelength calcium triplet lines.  We also created a linear calibration based on our stellar cluster (GC) calibrators (NGC 6791, Pal 14, NGC 7078) to check that there were no strong dependencies on instrument resolution.  

\begin{figure}
\begin{center}
\ifpdf
\includegraphics[width=0.94\textwidth]{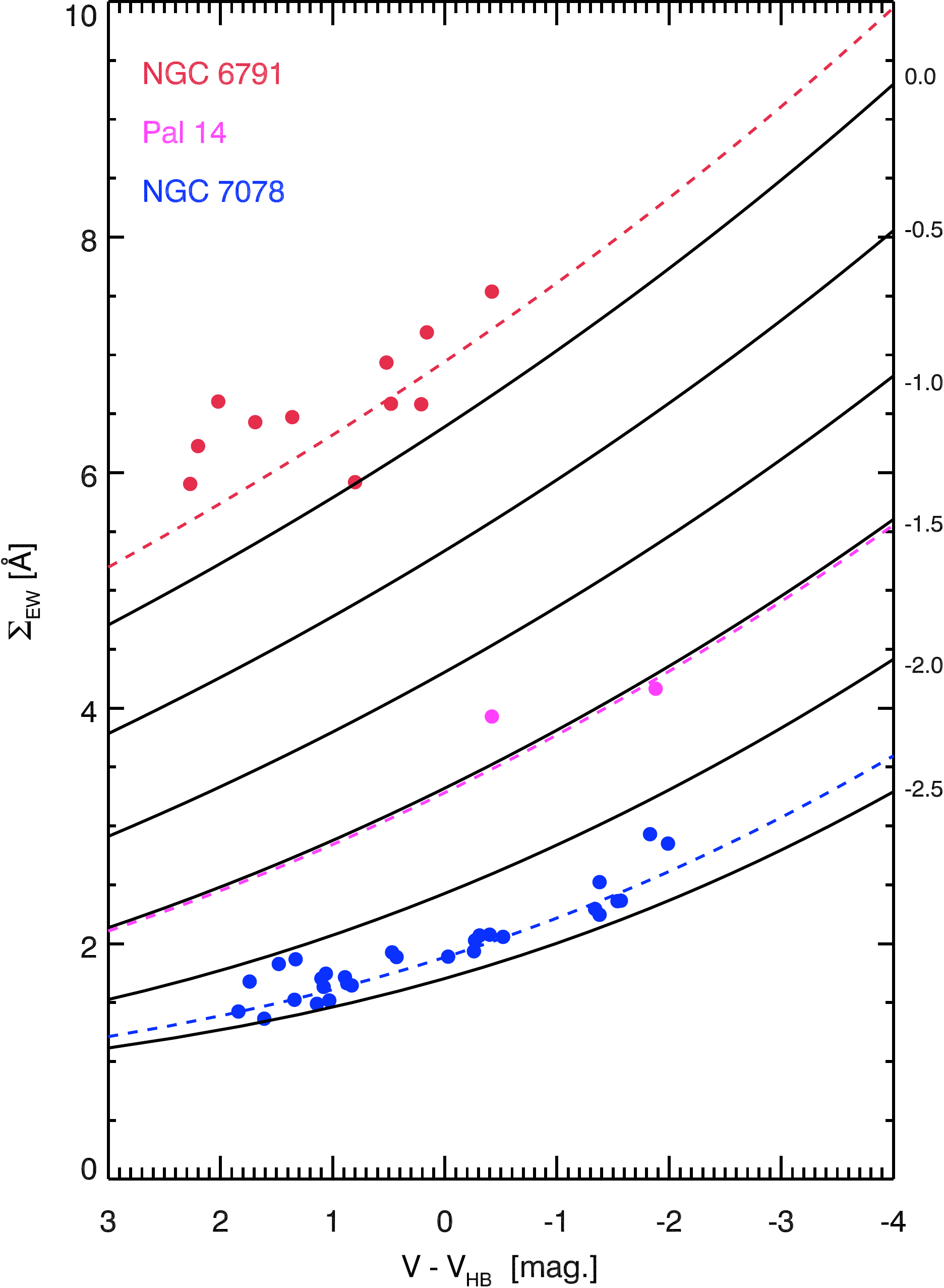}
\else
\includegraphics[width=0.75\textwidth]{ewvhb.eps}
\fi
\caption{Equivalent width as a function of magnitude relative to the horizontal branch for stars in the calibrating clusters NGC 6791 (\emph{red}), Pal 14 (\emph{magenta}), NGC 7078 (\emph{blue}).  Shown as the solid black lines is the [Fe/H] calibration of \cite{Starkenburg10}, illustrating the non-linearity clearly.  Calibration lines proceed in constant [Fe/H] values from solar to $-2.5$ dex in 0.5 dex steps, according to the given calibration.  Dashed red, magenta and blue lines are [Fe/H] values corresponding to the calibrating cluster's mean metallicity as taken from \cite{Harris96} and \cite{Dotter08} (Pal 14).}
\label{fig:ewcal}
\end{center}
\end{figure}

Figure \ref{fig:ewcal} shows the summed equivalent width as a function of V magnitude relative to the Horizontal Branch for the calibration based on Equation 5 from \cite{Starkenburg10}.  The summed EWs of the DEIMOS stars track the \cite{Starkenburg10} relations closely, perhaps suggesting that at least for moderate resolution spectra, pixel integration of the CaT lines produces EWs which fall on the \cite{Starkenburg10} relations.  The non-linearity in the low EW and faint end of the parameter space is evident in this diagram.  While the calibrating Globular Cluster stars shown in that figure extend to low magnitudes below the horizontal branch, we note that most of our WLM member stars are in the region of the parameter space, where large differences between the linear and non-linear calibrations are not expected.

While there is a more recent [Fe/H] scale based on globular clusters from \cite{CG09}, it is an average of four past scales (including the \cite{CG97} scale).  Unfortunately only one of our calibrating clusters is directly measured in the \cite{CG09} sample (NGC 7078) for which the \cite{Starkenburg10} calibration shows excellent agreement.  From the FORS2 calibrating clusters alone we find that the difference using the \cite{Cole04} calibration between the \cite{CG97} and \cite{CG09} scales for NGC 104 and NGC 7078 is only $0.06$ and $0.20$ dex respectively.  Given the relative size of the random uncertainties, and other factors discussed in joining these merged datasets this choice of absolute [Fe/H] scales does not introduce a change in our analysis.

\begin{figure}
\begin{center}
\ifpdf
\includegraphics[width=0.84\textwidth]{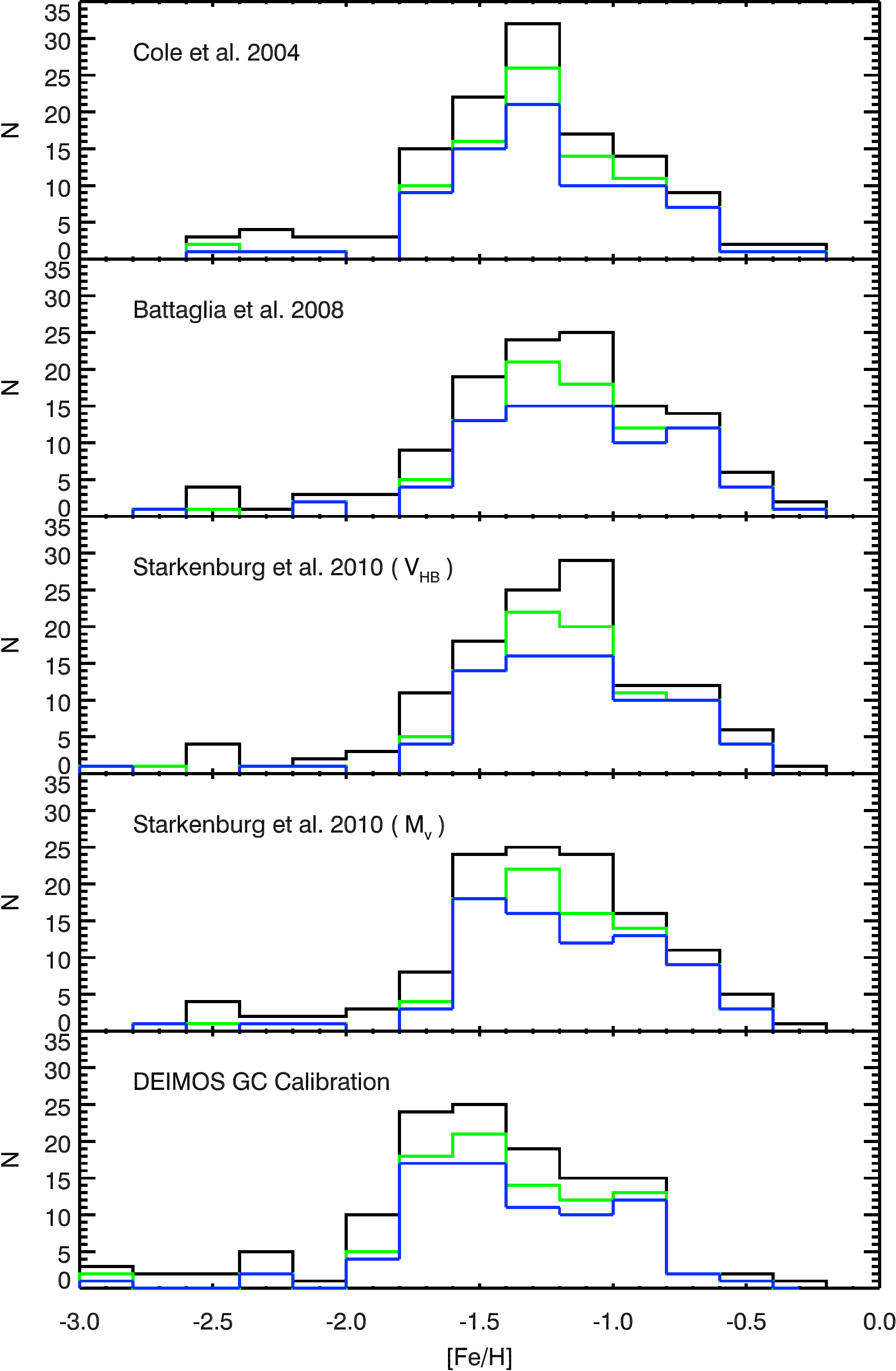}
\else
\includegraphics[width=0.72\textwidth]{catfeh5wlm.eps}
\fi
\caption{Full metallicity distribution functions for 126 stars out of 180 stars in our WLM dataset in which the spectra had S/N $\gtrsim 10 \rm{\AA}^{-1}$.  Each panel shows the distribution derived from a different empirical Calcium Triplet calibration. Full distribution is shown in black, original 78 stars from the FORS2 dataset of Paper I are shown in blue, and DEIMOS spectra of the highest S/N quality in green.  Within a given calibration, samples show good agreement, providing confidence in even the lowest S/N DEIMOS stars}
\label{fig:e5mdf}
\end{center}
\end{figure}

This is apparent from examining the metallicity distribution functions (MDFs) for 126 stars of the 180 member stars which had sufficiently high signal-to-noise ($\gtrsim10 \rm{\AA^{-1}}$) - shown for each of the five calibrations in Figure \ref{fig:e5mdf}.  While the calibrations are all in good agreement with one another for the relatively bright and metal rich stars of WLM, we wish to compare WLM to low metallicity dSph systems and therefore adopt the non-linear calibration of \cite{Starkenburg10} based on the horizontal branch magnitude for the analysis in the rest of this paper. This allows us to compare WLM to the dSphs on a consistent scale that does not suffer from saturation biases which would be prevalent in the faint, low metallicity dSph stars.

\begin{figure}
\begin{center}
\ifpdf
\includegraphics[width=0.98\textwidth]{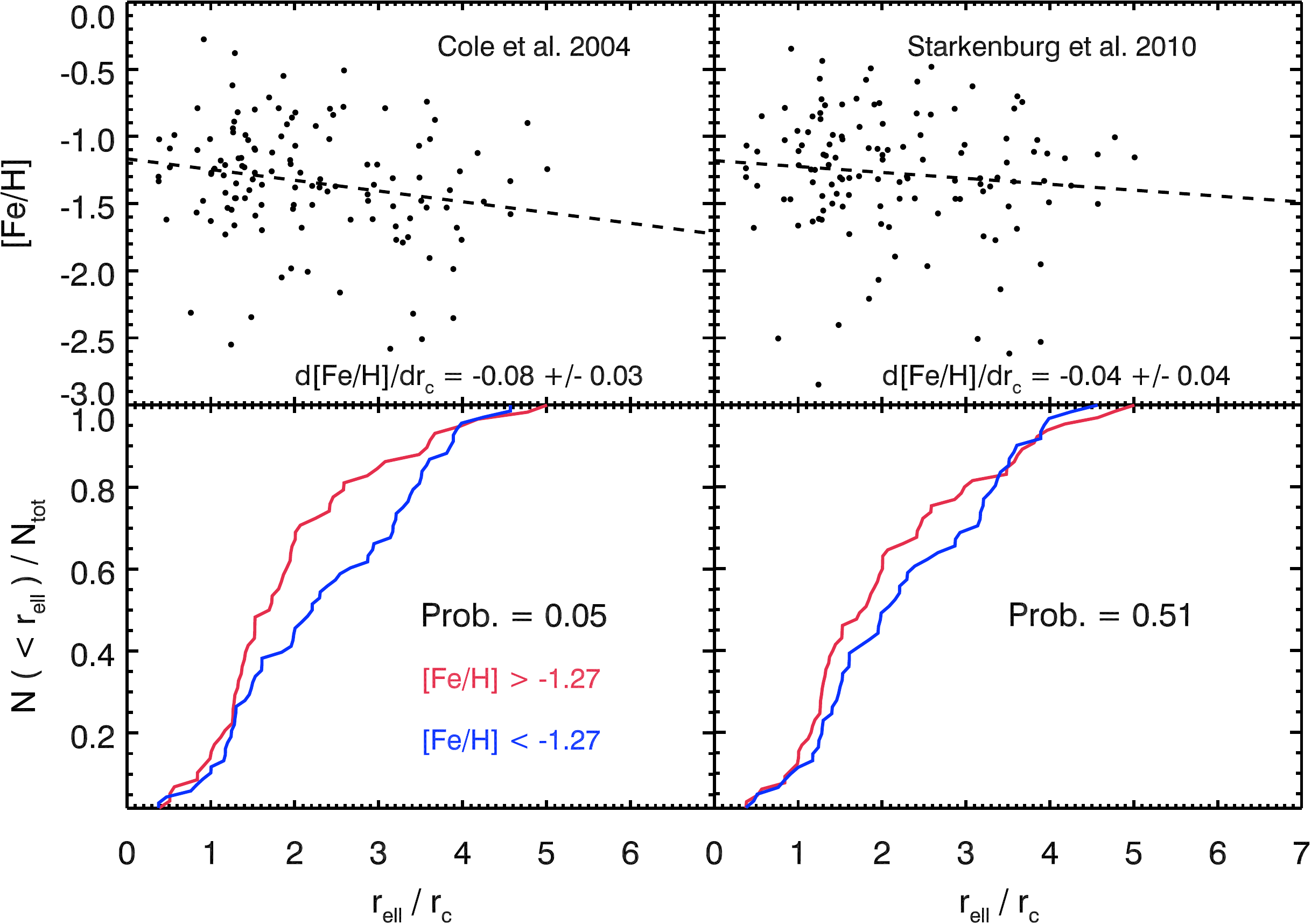}
\else
\includegraphics[angle=270,width=0.98\textwidth]{3rellcdff.eps}
\fi
\caption{Impact of [Fe/H] CaT calibration on the presence of spatially segregated subpopulations and gradients.  Left panel shows the linear calibration of \cite{Cole04} adopted in Paper I, compared to the non-linear calibration of \cite{Starkenburg10} in the right panel adopted for this work. Probability shown in the lower panels are computed from a two dimensional, two-sided Kolmogorov-Smirnov test, and represents likelihood that the metal poor and metal rich populations are drawn from the same parent distribution in each calibration case.}
\label{fig:rell3cdfwlm}
\end{center}
\end{figure}

Metallicity calibration biases are especially important to consider when studying the spatial distribution of metallicities within a galaxy. In Figure \ref{fig:rell3cdfwlm} we show the effect on spatial variations in [Fe/H] for two different CaT based [Fe/H] calibrations for WLM.  There are visible differences even for the relatively bright and metal rich stars in our sample, which would be even more severe in low metallicity systems.  As gradients have such large implications for dwarf galaxy formation and evolution, care must be taken during interpretation, and this starts with an accurate understanding of the calibration biases.  

Uncertainties on the metallicities were propagated from the initial line width measurements to the calibrated [Fe/H] values, as in Paper I.  The mean uncertainty of the FORS2 stars is 0.25 dex, and 0.26 dex for the DEIMOS stars.  While higher resolution, some of the DEIMOS data are much lower signal to noise, which allowed us to derive reliable metallicity estimates for only $50\%$ of those spectra.

\subsubsection{Consistency Checks}
The MDFs presented in Figure \ref{fig:e5mdf} show good qualitative agreement and have similar dispersions and mean values within the uncertainties.  While ideally a homogeneous instrument setup for the full sample of stars is preferred, we find consistent metallicity signatures between Paper I and this joint sample within the large uncertainties.  Evidence for this comes from repeat measurements of the two stars in common between the DEIMOS and FORS2 observations (which show a difference of $0.13 \leq \Delta[Fe/H] \leq 0.31$), and a similar mean metallicity ($-1.27 \pm 0.04$; $-1.28 \pm 0.02$) for all of the stars in Paper I and this full sample.  Additionally the radial gradient computed in Paper I ($\rm{d[Fe/H]/d{r_{c}}} \simeq -0.076 \pm 0.03$ dex r$_{\rm{c}}^{-1}$) is unchanged when we apply the \cite{Cole04} calibration to the joint sample of FORS2 and DEIMOS spectra ($\rm{d[Fe/H]/d{r_{c}}} \simeq -0.08 \pm 0.03$ dex r$_{\rm{c}}^{-1}$).  Therefore, while there are subjective choices on the metallicity scale, EW measurements, and CaT calibration, these changes are all within the errors.  This suggests that given the large uncertainties inherent in spectroscopy of stars in distant galaxies like WLM, our joint MDF is sufficiently robust to analyze the global metallicity properties
\\

\subsection{Age Derivations}
Age derivations were discussed in Paper I and Paper II for the WLM sample.  Ages were derived using the published photometry and the \cite{Demarque04} stellar evolution models.  The older library is chosen due to the metallicities in the larger sample and new calibration being outside the range of the Victoria-Regina models used in Paper I.  In addition a greater flexibility in $\alpha-$abundances is possible with the \cite{Demarque04} models.  The V and I photometry, reddening, and distance moduli were taken from Papers I and II on WLM, and discussed therein.  The ages were interpolated using the grid of isochrones, the dereddened photometry, and spectroscopic [Fe/H] and [$\alpha$/Fe] estimates.  As in Paper II the [$\alpha$/Fe] values were interpolated as a function of [Fe/H] using the literature values from \cite{Colucci10,Venn03,Bresolin06} to describe the mean trend of [$\alpha$/Fe] versus [Fe/H] in WLM.  Errors were assigned by propagating the photometric, reddening and distance modulus uncertainties, as well as the spectroscopic abundance uncertainties into the position of the star on the colour magnitude diagram.  Ages derived using this method will be valid in a differential sense within a sample, as there are strong systematic uncertainties between the stellar evolution libraries used in various studies.  However, the metallicity uncertainties dominate over the choice of evolution library for such distant systems as WLM, therefore the general age-metallicity properties of WLM may be extracted.  The relative random uncertainty on age for an individual WLM star is $\sim 50\%$.

\subsubsection{Quantifying Systematic Age Errors}
The random error captures the uncertainty in derived age due to errors in colour, magnitude and [Fe/H], however there are three additional systematic errors not included in the previous section that must be quantified - AGB contamination, differential (internal) reddening, and variations in [$\alpha$/Fe].  WLM exhibits an extended SFH \citep{Dolphin00}, therefore it is highly probable to sample stars on the giant branch with ages $1.6-12$ Gyr.  In addition the distance of WLM makes it difficult to accurately differentiate second ascent giant branch stars with much confidence from photometry.  This means that within the sample there may be AGB stars; these do not affect the derived [Fe/H] or velocities but can produce a bias in the inferred age.  Using the SFH of \cite{Dolphin00}, it is possible to roughly estimate the AGB contamination rate within the colour and magnitude range of the WLM spectroscopic targets.   A conservative upper limit on the contamination fraction is $\sim 40\%$, with about 1/3 of those AGB stars being younger than 2.5 Gyr, and a third older than 9 Gyr.  Using a grid of isochrones, it is possible to work out for a given colour and magnitude the difference in age between an RGB and AGB star.  The systematic age error due to AGB contamination is found to be strongest at young ages.  Specifically, an AGB star of 1.6 Gyr would have its age underestimated by $20\%$ if it were considered an RGB star in the sample.  This percentage drops to $10\%$ for a 2 Gyr star, and $5\%$ for a 10 Gyr star.

The unknown nature of differential internal reddening, and star-to-star variations in $\alpha-$element abundances in WLM stars will contribute additional systematic errors.  To numerically estimate the \emph{combined} systematic uncertainty due to these two factors and the above mentioned AGB contamination, we proceeded as follows.  For a test star of a given true age and [Fe/H], and fixed evolutionary position $\sim 0.5$ magnitude below the tip of the RGB, we drew randomly a possible variation in (V-I) (due to reddening), and [$\alpha$/Fe], as well as gave it a $50\%$ chance at being an AGB star.  The distribution of internal reddenings was drawn from a Gaussian of $\sigma_{V-I} = 0.03$ mag (for comparison, the line of sight reddening in the direction of WLM is E(V-I)$ = 0.037$ \citealt{SFD98}).  The distribution of differential [$\alpha$/Fe] was taken from a Gaussian of $\sigma_{\alpha} = 0.05$ dex - which was chosen primarily to keep [$\alpha$/Fe] within the range of the isochrone grid.  In a given iteration, the test star had its age rederived using the new colour and magnitude on a grid of isochrones reflecting its new [$\alpha$/Fe].  If the star was also drawn to be an AGB star, the new age was modified by the systematic age offsets discussed in the above paragraph.  

\begin{figure}
\begin{center}
\ifpdf
\includegraphics[width=0.98\textwidth]{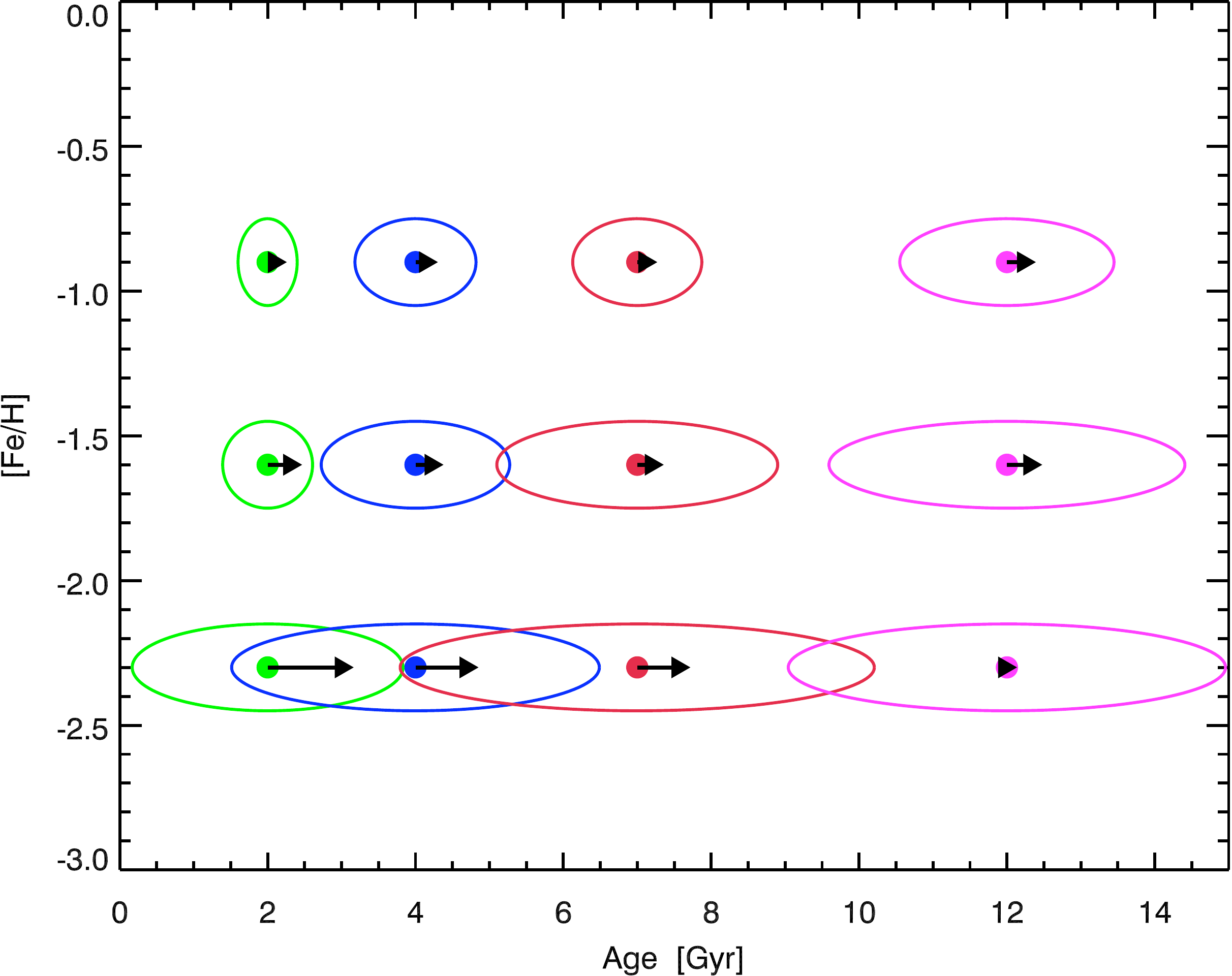}
\else
\includegraphics[angle=270,width=0.98\textwidth]{synthagen}
\fi
\caption{Representation of systematic age errors on artificial input stars of various [Fe/H] and true age.  Each oval shows an estimate of the combined systematic uncertainty in recovered age due to the effects of differential reddening, [$\alpha$/Fe] variations, and AGB contamination for 10000 artificial star tests.  Arrows show movement in the mean recovered age from the combined effects.}
\label{fig:synthage}
\end{center}
\end{figure}

In Figure \ref{fig:synthage} we show the systematic errors due to these combined effects for 10000 iterations on each input star of a given canonical age and [Fe/H].  While AGB contamination most strongly impacts the age systematics at young ages, the effect of differential reddening and $\alpha-$element variation dominates for low metallicity stars.  The standard deviation of Monte Carlo trials are indicated as ellipses, with the mean movement indicated by the black arrows.  Where appropriate in this work and in Paper II, we adopt the semi-major axis of the ellipses as an estimate of the \emph{total systematic} uncertainty for stars of various ages and metallicities, and incorporate that along with the individual random uncertainty on age for a star.  The distance of WLM makes deriving ages difficult, however we are aided by the relatively metal rich and young populations of stars in this sample.

\section{Results}
We now compare the distribution of metallicities, both spatially and with age (where possible), using the spatially extended sample of metallicities for hundreds of stars in each of WLM, LMC, SMC, Fornax, Leo I, Sculptor, Leo II, Sextans, and Carina.  Through this analysis we will test whether an isolated galaxy such as WLM, with its relatively quiescent tidal evolution and gas content history, shows differences in chemical evolution compared to the gas poor-tidally disturbed dSphs or gas rich tidally disturbed Magellanic Clouds\footnote{It should be kept in mind that tidal influences on the Magellanic Clouds may have only begun recently ($\leq 2$ Gyr) if they are on first infall to the MW \citep{Besla07,BK11}.}.

\subsection{Metallicity Distribution Functions}
In comparing the metallicity distribution functions (MDFs) between the dwarf galaxies, it is important to be aware of the sample sizes and spatial extents of the datasets for comparison galaxies, and where [Fe/H] has been calculated using differing methods. Both the K10 and the DART sample have [Fe/H] estimates that should be free of calibration biases (see $\S 3$).  There remains systematic differences due to the varying spatial coverage and number of stars in each study however.  The global metallicity distribution may be biased in studies where only central regions of a galaxy are sampled, especially in the case where there is a radial metallicity gradient.  Similarly when the sample size is small the metallicity may be underestimated, as a minority metal poor population is more difficult to sample as efficiently due to a population bias resulting from mass dependent stellar evolutionary timescales\footnote{For a given time interval, more young, high mass, high metallicity stars will evolve to the TRGB than older, low mass, metal poor stars in the same time interval.  The dominant fraction of young stars at the TRGB is large enough to outnumber the metal poor old stars, despite the contradictory initial relative numbers of the IMF. cf., \citealt{Cole08}}  This effect will be enhanced in the presence of any age-metallicity relationship and is still present for populations with extended SFHs like WLM.

\begin{figure}
\begin{center}
\ifpdf
\includegraphics[width=0.98\textwidth]{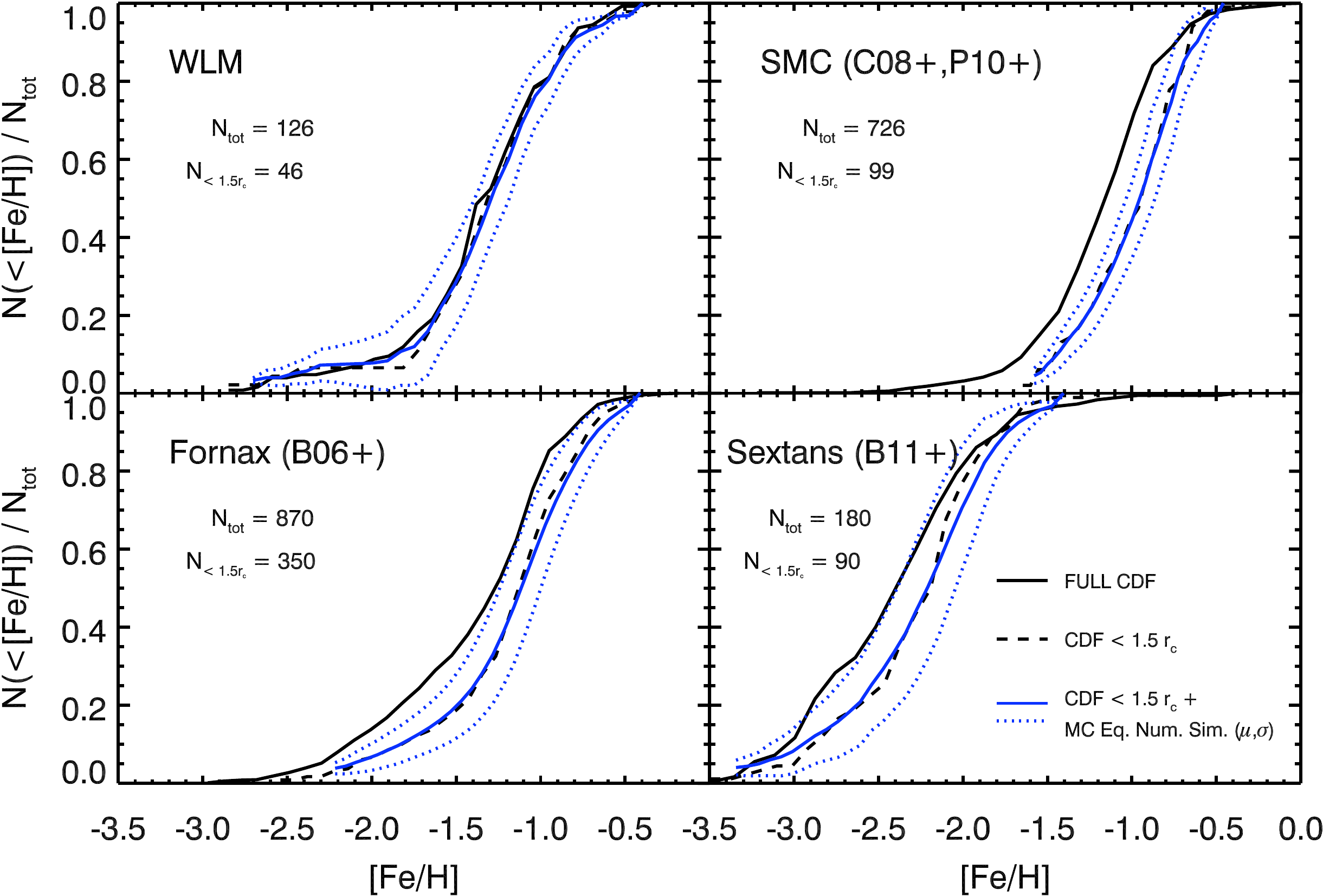}
\else
\includegraphics[angle=270,width=0.98\textwidth]{fehcutcomp.eps}
\fi
\caption{Panels show example of spatial and number biases in metallicity distributions of four dwarf galaxies in the sample.  In addition to the full CDFs, the CDF within the inner $1.5 \rm{r_{c}}$ is shown (\emph{dashed line}), along with the mean and dispersion from 10000 iterations of random sampling of 31 stars from that inner region in each galaxy (\emph{blue solid and dashed}).  The small difference between the solid blue and black dashed lines indicates that the primary bias in comparing samples is consistent spatial coverage, especially when dealing with galaxies that show strong metallicity gradients.}
\label{fig:fehcutcomp}
\end{center}
\end{figure}

To study the impact of these biases, MDFs were computed for the galaxies using an equal spatial range, and with equal numbers of stars.  The samples were restricted in these test cases to stars within $1.5r_{c}$ of the dwarf galaxy center (roughly the smallest radial extent of dSph samples in the K10 catalogue).   Any one galaxy that has a strong metallicity gradient will be biased in this case, however the MDFs of the inner regions will be much more appropriate to compare to each other in a differential sense.  To study the bias in sample size, the datasets were also resampled from the inner $1.5r_{c}$ so that they have equal numbers of stars.  In this case the sample sizes have been constructed to have 31 stars (equal to the smallest sample of stars within $1.5r_{c}$ in any of the galaxies).  Each of the galaxies were resampled from the larger population 10000 times, computing a cumulative metallicity distribution function (CDF) each time.  An average CDF was constructed based on 31 stars in each galaxy.  Figure \ref{fig:fehcutcomp} shows an example of the mean (solid blue) and dispersion (dotted blue line) of resampled CDFs for several of the galaxies in this study, as well as the impact of the radial cutoff.  This shows that the primary bias affecting comparison of different studies is their spatial extent, rather than any sampling bias from differing numbers of stars.  We have therefore selected surveys which have resolved spectroscopic data out to at least $75\%$ of the galaxy's tidal radius in hopes of minimizing any spatial bias when comparing the dwarf galaxies, as shown in Table 1.

\begin{figure}
\begin{center}
\ifpdf
\includegraphics[width=0.85\textwidth]{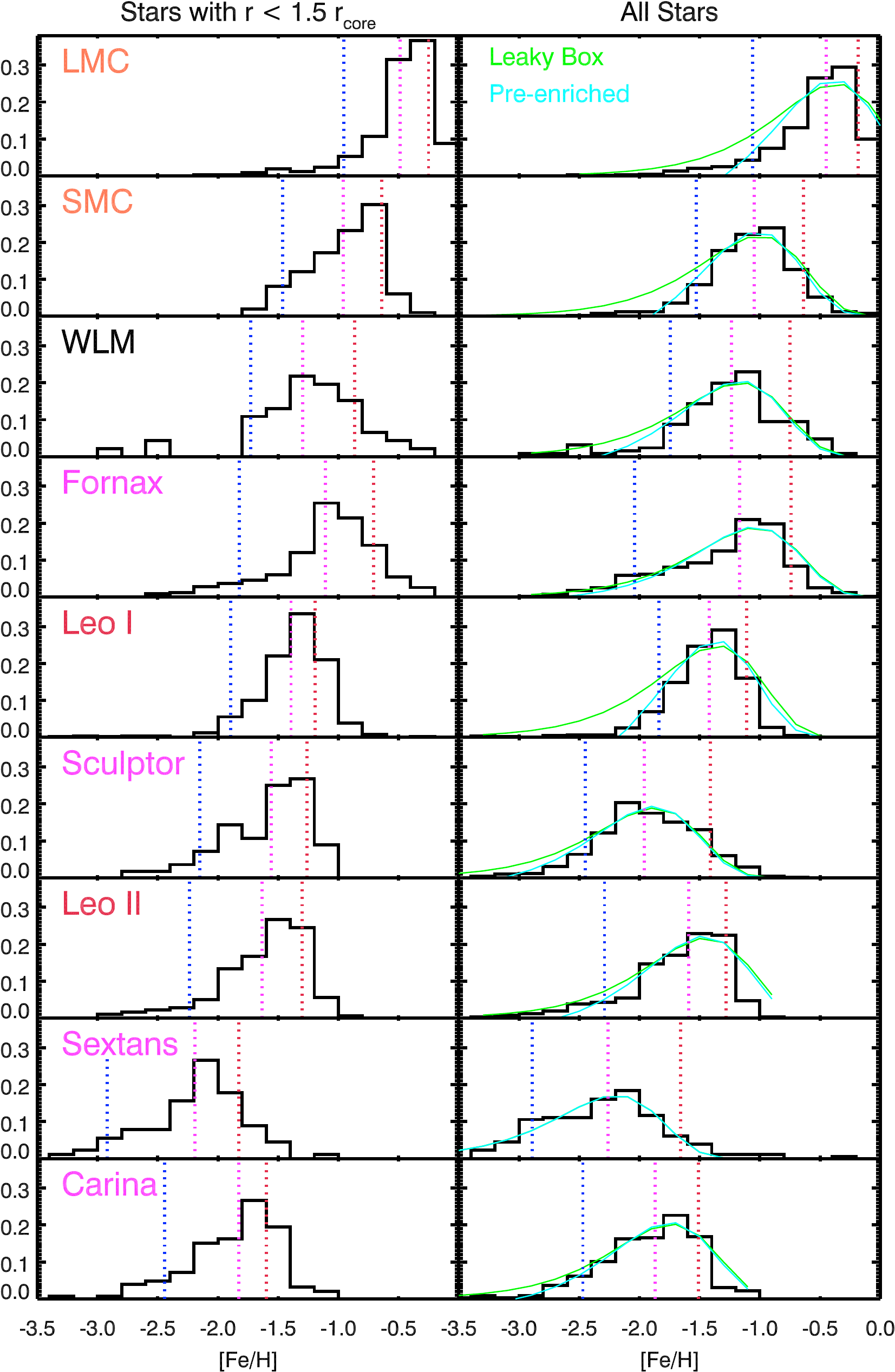}
\else
\includegraphics[width=0.55\textwidth]{difcomp.eps}
\fi
\caption{Differential MDFs for the Local Group dSphs and dIrrs considered in this paper, ordered by decreasing luminosity from top to bottom.  Galaxy names are colour coded by the first author of the samples: red (Kirby), magenta (DART survey), orange (Magellanic Cloud surveys) (see Table 1). Left panel shows metallicity distributions in each galaxy where only stars within $1.5 r_{c}$ have been considered, right panels show full sample of stars for a given galaxy. Panels show normalized histogram only with respect to the particular sample (i.e., fraction of total stars for the right hand panels, or fraction of stars within $1.5 r_{c}$ for the left panels).  Dotted blue, magenta, and red lines indicate the 10th, 50th, and 90th percentile [Fe/H] values for each sample.  Simple leaky box and pre-enriched chemical evolution models \citep{Prantzos08} are overlaid on the full samples in green and cyan.}
\label{fig:difcomp}
\end{center}
\end{figure}

Figure \ref{fig:difcomp} shows the MDF of WLM, the Magellanic Clouds and the six dSphs, ordered by luminosity. The full sample and central regions are shown for each galaxy.  In each panel the 10th, 50th, and 90th percentile metallicities of the MDFs are also shown (as dotted lines).  The mean (median) metallicity of WLM is [Fe/H]$ = -1.28 (-1.24) \pm 0.02$ dex, in good agreement with the trend shown by the Local Group luminosity-metallicity relation (cf., \citealt{Woo08}).  The uncertainty on the value for WLM represents the error on the mean, however typical systematic uncertainties for both synthesis and equivalent width based measurements from the literature are $\sim 0.15$ dex.  The median value of WLM lies within the range of [Fe/H] distributions shown by the dSphs, and is in close agreement to the more luminous members, Fornax and Leo I as well as the SMC.  The metal poor population below [Fe/H]$ = -1.74$ forms a $\sim 10\%$ minority population in WLM, and stars are found as metal poor as [Fe/H]$ = -2.85$ dex.  RGB stars with metallicities as enriched as [Fe/H]$ = -0.35$ dex are found, consistent with the results from \cite{Venn03} for two supergiants in WLM.  Table 2 tabulates the percentiles of the MDF for each galaxy. The mean and extrema metallicities of the isolated, gas rich WLM dIrr show little differences from the luminous dSphs, nor the gas rich but (recently) tidally perturbed SMC.  

WLM shows a similar median metallicity to some classical dSphs and the SMC, however there appears to be a slight preference for the lower mass dSphs to show more asymmetric, extended metal poor tails relative to their mean metallicity.  Fornax, Leo II, Sextans, and Carina all show a significant asymmetry to their MDF compared to the dIrrs - however it may be that the metal poor stars in the gas rich dwarfs are simply more difficult to sample efficiently in such cases with ongoing star formation (see $\S4.1$).  The difference in metal poor tails can be seen qualitatively in the right panels of Figure \ref{fig:difcomp}, where leaky box and pre-enriched chemical evolution models have been overlaid.  The formalism of \cite{Prantzos08} for the two models has been used, and the best fitting models determined through a maximum likelihood approach assuming Poisson errors on the distributions.  The estimated yields and initial metallicity for both chemical evolution models are listed for each galaxy in Table 2.  Qualitatively the simple leaky box models provide reasonable fits to many of the dSphs, but the dIrrs and Leo I appear more closely fit by the pre-enriched solutions (see also \citealt{Gullieuszik09}).  In the case of Leo I this could be due to the low metallicity stars lying at large radii (see $\S 4.2$) outside the spatial coverage of the K10 survey - as the DART sample of Fornax (which is nearly the same luminosity as Leo I) samples out twice as far in radius and shows a significant metal poor tail.  Alternatively for Leo I, \cite{Lanfranchi10} showed that models with infalling pristine gas could also reproduce the MDF.

\begin{figure}
\begin{center}
\ifpdf
\includegraphics[width=0.8\textwidth]{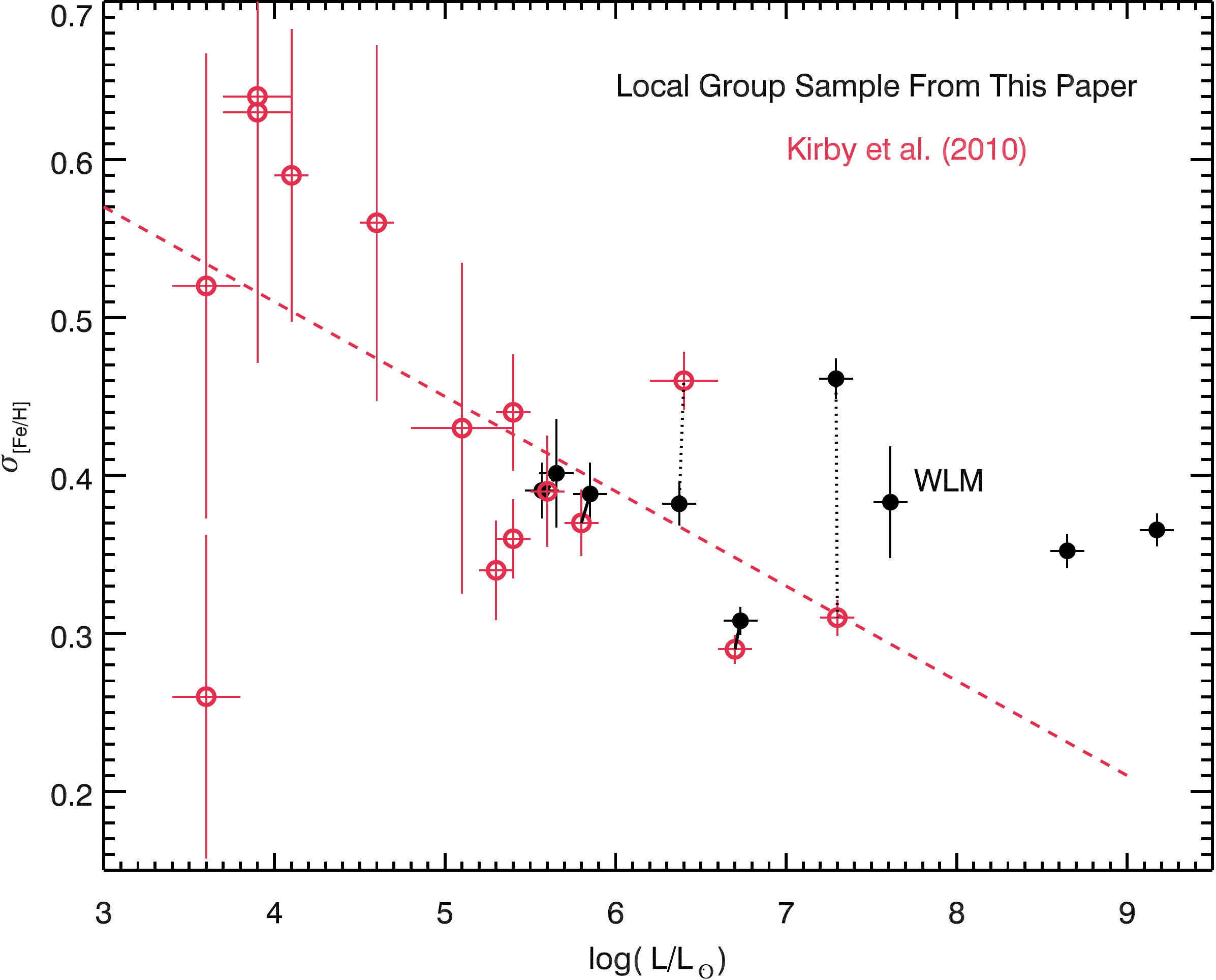}
\else
\includegraphics[angle=270,width=0.8\textwidth]{siglumsf.eps}
\fi
\caption{Intrinsic dispersion in [Fe/H] as a function of host galaxy luminosity for the sample of \cite{Kirby10} (\emph{red circles}) and other dwarf galaxies considered in this paper (\emph{filled black dots}).  Solid lines connect values where the intrinsic dispersion has been measured on the same data in this work, and in \cite{Kirby10} as a check on consistent methodology for removing the contribution of measurement uncertainty to the dispersion.  Dotted lines connect the derived dispersion between the different DART and Kirby et al. samples for Fornax, Sculptor and Sextans.  Red dashed line shows the best fitting relation found by \cite{Kirby10} including the ultra faint dSphs.}
\label{fig:siglums}
\end{center}
\end{figure}

The spread in [Fe/H] of a dwarf galaxy has been found to anticorrelate with mean metallicity in the recent study of \cite{Kirby11a}.    We revisit this relation here, as our literature sample explores both higher luminosity galaxies (some of which are gas rich, and one of which is isolated), and typically have larger spatial coverage.  The intrinsic dispersions have been calculated by subtracting in quadrature the mean error in metallicity for a sample from the total measured dispersion.  Figure \ref{fig:siglums} shows the intrinsic spread in metallicity, $\sigma_{\rm{[Fe/H]}}$, for the data in this paper as well as those derived in \cite{Kirby11a} and the linear correlation those authors found.  The uncertainty on the dispersions are calculated using relations from $\S 3.1$ of \cite{Hargreaves94} and the intrinsic and raw dispersions from this work and K10. Figure \ref{fig:siglums} suggests that the K10 relation still provides a good description of the large spread seen in [Fe/H] for the Ultra Faint Dwarfs (although Bootes I was recently found to lie off this relation; \citealt{Lai11}). However at high luminosities ($\gtrsim 10^{5} $L$_{\odot}$) the dispersion in metallicity may saturate, as hinted at in \cite{Norris10}, as it shows little change even up to the brighter dIrrs such as WLM and the Magellanic Clouds.  This may be due to the high luminosity systems having an enrichment timescale that is similar, or above some threshold duration to produce a well mixed ISM (see $\S 5.1$; also \citealt{Leaman12b}).  Alternatively it could suggest that the lowest mass systems are more impacted by gas and metal outflows, driving them to larger metallicity dispersions, compared to the higher mass galaxies which retain a larger fraction of their gas and metals.

\subsection{Spatial Variations in Chemistry}
Paper I showed that the metal rich and metal poor stars in WLM are similar in their spatial distributions with slightly more metal rich stars in the inner regions.  With only 78 stars in the original WLM sample however, it was difficult to interpret the change as due to a superposition of two populations, or a smooth gradient.  Here we analyze the spatial distribution of the WLM stellar populations using the new DEIMOS data and CaT-[Fe/H] calibration, and examine the metallicity gradient in comparison to those of dSphs (i.e., \citealt{Tolstoy04,Battaglia06}).  With the new stars and calibration by \cite{Starkenburg10}, the metallicity distribution has changed and the spatial signatures are altered from Paper I.  Figure \ref{fig:rell3cdfwlm} shows that the new calibration clearly favours a milder gradient.

\begin{figure}
\begin{center}
\ifpdf
\includegraphics[width=0.90\textwidth]{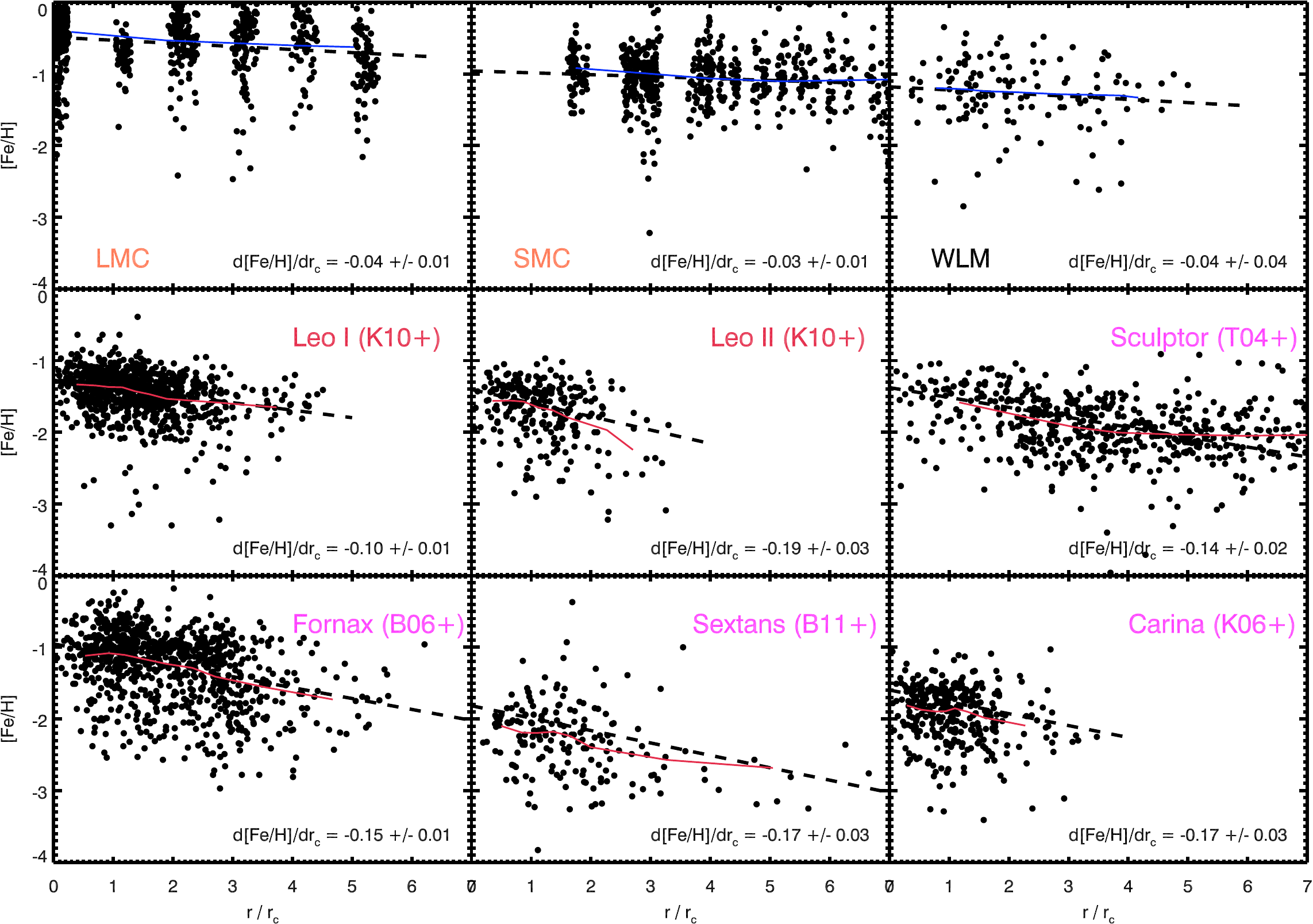}
\else
\includegraphics[angle=270,width=0.90\textwidth]{allrellfeh2n.eps}
\fi
\caption{Plot of [Fe/H] vs. geometrical radii in units of core radii for WLM and the Local Group dwarfs from literature. Galaxy names are colour coded by the first author of the samples: red (Kirby), magenta (DART survey), orange (Magellanic Cloud surveys) (see Table 1). Dashed lines show weighted linear fits to the data.  Solid lines show the running boxcar averages of Figure \ref{fig:rellover} overlaid.  The rotating dIrrs appear to show statistically flatter abundance gradients than the dispersion dominated dSphs (weighted averages of $-0.03\pm0.01$ vs. $-0.13\pm0.01$).}
\label{fig:rellfeh}
\end{center}
\end{figure}

Metallicity gradients in dSphs are still common to varying degrees as shown in \cite{Kirby11a}.  However the spatial coverage in that work is relatively limited compared to the samples from \cite{Walker09b}, or the DART survey.  Here we can explore comparisons of [Fe/H] gradients in a rigorous manner for the Local Group galaxies due to the large spatial coverage ($r_{max} \geq 0.75r_{t}$) in the literature sample.  Figure \ref{fig:rellfeh} shows [Fe/H] as a function of elliptical radius (in units of core radii), with the dashed lines showing linear least squares weighted fits to the data.  

WLM shows a mild gradient with $\rm{d[Fe/H]/dr_{c}} = -0.04 \pm 0.04$ \rm{dex r$_{\rm{c}}^{-1}$}, similar to the Magellanic Clouds.  It should be noted that the LMC RGB metallicities of \cite{Carrera08b} show a correlation with magnitude, which manifests itself as a systematic steepening of the gradient.  Correction of this reduces the measured gradient by half, consistent with an extended sample of RGB stars from a forthcoming study ($\rm{d[Fe/H]/dr_{c}} = -0.008 \pm 0.004$ \rm{dex r$_{\rm{c}}^{-1}$}; Cole et al., in prep.) - therefore the LMC gradient shown here could be taken as a lower limit.  In contrast the dSphs show radial metallicity profiles which are much steeper.

\begin{figure}
\begin{center}
\ifpdf
\includegraphics[width=0.9\textwidth]{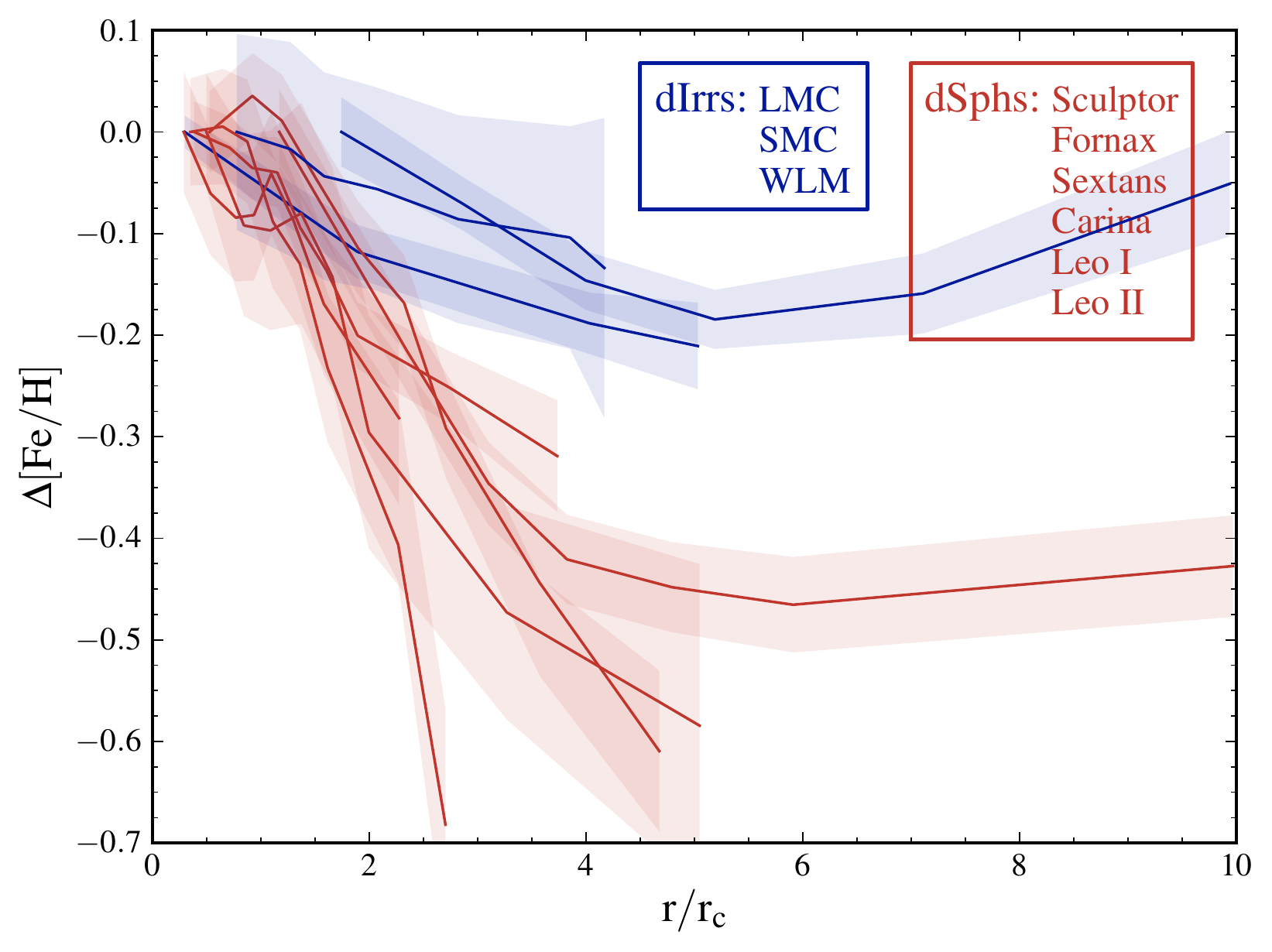}
\else
\includegraphics[width=0.9\textwidth]{delfeh3.eps}
\fi
\caption{Running boxcar averages of [Fe/H] as a function of radius in dIrrs (\emph{blue}), and dSphs (\emph{red}).  Radial profiles have been smoothed by a factor of 2 for clarity, and normalized by the metallicity of the central regions of each dwarf galaxy.  The blue and red shaded bands represent the associated $1-\sigma$ uncertainties on the running averages.}
\label{fig:rellover}
\end{center}
\end{figure}

Many of the individual galaxies in Figure \ref{fig:rellfeh} show complex trends of [Fe/H] with radius, and as noted in \cite{Battaglia11} some galaxies show steep profiles in the inner few core radii, which then flatten to a low metallicity plateau.  In such cases the gradient may be a superposition of two populations with differing concentrations - a possibility that is difficult to rule out quantitatively.  We have computed running boxcar averages of [Fe/H] versus radius in Figure \ref{fig:rellover} to track the mean metallicity more precisely.  The dIrrs still show statistically flatter radial [Fe/H] gradients relative to the dSphs (especially considering the uncorrected systematic with the LMC profile).  The low metallicity plateau in Sculptor is quite evident at large radii, and while there is scatter between any given dSph, overall they show much steeper metallicity drop offs in the inner three core radii.  Figure \ref{fig:rellover} clearly shows the necessity for very spatially extended spectroscopic surveys in order calculate the complete abundance gradient and global metallicity distribution, as within the inner $1.5r_{c}$ of the dSphs there is a stochastic behaviour before the abundances coherently fall at larger radii.

\subsection{Age Metallicity Relations}
\begin{figure}
\begin{center}
\ifpdf
\includegraphics[width=0.85\textwidth]{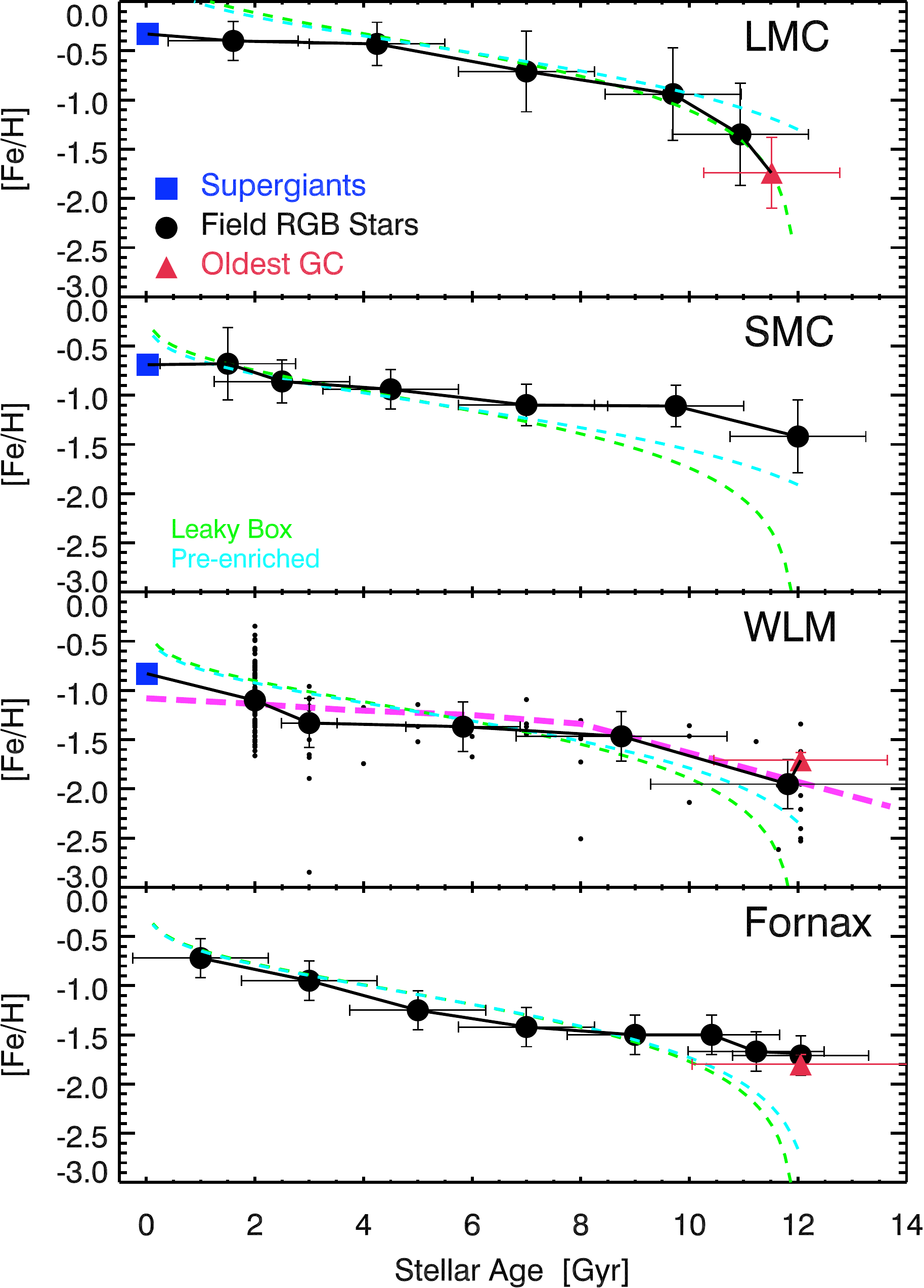}
\else
\includegraphics[width=0.75\textwidth]{wlmamrt3n.eps}
\fi
\caption{Age-metallicity relations for the LMC \citep{Cole04}, SMC \citep{Carrera08}, Fornax \citep{Battaglia06}, and WLM.  Binned RGB field stars are shown as large black circles for all galaxies, and in WLM the individual stars are shown as the small black dots.  The oldest cluster in the LMC and WLM \citep{Hodge99} are shown as the red triangles, and the A or B supergiant values as the blue squares.  Overlaid in green and cyan are the simple leaky box and pre-enriched chemical evolution models from Figure \ref{fig:difcomp}.  The HST SFH history solutions from the work of \cite{Dolphin00} are overlaid as the magenta line for WLM.}
\label{fig:wlmamr}
\end{center}
\end{figure}

Photometric SF histories already suggest that WLM experienced extended star formation over its lifetime \citep{Dolphin00}, and with spectroscopic [Fe/H] and age estimates in WLM, we can compare the age metallicity relation to several other Local Group dwarfs in a differential sense.  Figure \ref{fig:wlmamr} plots the age metallicity relations (AMRs) for the three dwarf irregulars as well as Fornax.  For the LMC and SMC we have used the published age-metallicity relations in \cite{Cole04} and \cite{Carrera08} respectively.  Similarly with Fornax we have adopted the published mean age-metallicity relation from \cite{Battaglia06}.  In those three studies, stars with ages older than 10 Gyrs have been scaled as $t_{new} = 10 + 0.41(t - 10)$, in order to homogenize the oldest stars to a consistent maximum age, as the studies differ in the uppermost age the stellar evolution libraries consider.  For WLM five bins have been computed for the AMR, and the error on the mean for the metallicity, and combined random and systematic age errors computed for each bin ($\S 3.3.1$).  Where applicable the field RGB star data are supplemented with the [Fe/H] values from supergiant studies by \cite{Bresolin06}, \cite{Venn03} and \cite{Levesque06}.  Similarly the oldest globular clusters in the LMC \citep{Colucci11}, Fornax \citep{Strader03} and the one GC in  WLM \citep{Colucci10} were added and metallicities taken from those spectroscopic studies.  There appears to be good agreement between these anchor points and the youngest and oldest field RGB star age-metallicity data.  The simple leaky-box and pre-enriched chemical evolution models which were fit to the MDFs in Figure \ref{fig:difcomp} have been overlaid as well.  

The AMRs of SMC, Fornax, and WLM in Figure \ref{fig:wlmamr} are all qualitatively similar in their shape, with relatively shallow metallicity enrichment over intermediate ages, and only an offset in metallicity at each age.  WLM shows slightly more rapid metal enrichment in the last 3 Gyr compared to any of the other galaxies, consistent with the burst of central SF found by \cite{Dolphin00}.  The nearly flat [Fe/H] values from $3-9$ Gyrs in WLM are quite similar to the SMC for that time frame, but Fornax shows a stronger increase in metallicity over the same timescale.  This is likely due to the strong burst of SF estimated to have occurred between 3-8 Gyr ago in Fornax based on its CMD analysis \citep{Stetson98,Saviane00,Tolstoy01,Coleman08,deBoer12}.  For the first two Gyr of it's lifetime, WLM does not appear to have undergone as rapid enrichment as the LMC.  The mean metallicity of WLM changes by $\lesssim 0.5$ dex in this period, similar to the SMC and Fornax, while the LMC changes its mean metallicity by nearly 1 dex.

The AMR of the LMC and SMC has been carefully examined by \cite{PG98}, where the primary difference in their AMR shapes was attributed to the relative strength of the SF burst(s) in the LMC.  WLM's SFH has been interpreted from HST colour magnitude analysis by \cite{Dolphin00}, and the computed AMR from that work is also shown overlaid on the WLM data in Figure \ref{fig:wlmamr}.  There is remarkable agreement between the shape of the AMR derived from our spectroscopic sample, and that inferred from the photometric study of \cite{Dolphin00}.  The similarities in the AMR of WLM to that of the dSph Fornax, illustrate that an isolated dIrr like WLM (which has experienced much less tidal perturbations over its lifetime, and exhibits different kinematics) has had chemical enrichment proceed similarly to both gas rich but tidally perturbed (SMC), and gas poor, low angular momentum (Fornax) systems of similar luminosity.

The dispersion in metallicity for stars of differing ages can provide insight into such processes as radial migration.  For the youngest stars in WLM ($\leq 2$ Gyr) the intrinsic dispersion in [Fe/H] is found to be $\sigma_{1-2} = 0.15 \pm 0.04$ dex.  This increases to $\sigma_{3-9} = 0.32 \pm 0.07$ dex for stars $3-9$ Gyr, and $\sigma_{10-13} = 0.33 \pm 0.09$ dex for stars $\geq 10$ Gyr.  Using the sample of \cite{Cole05} and identical age cuts for the LMC, the intrinsic dispersion in [Fe/H] is $\sigma_{1-2} = 0.13 \pm 0.01$, $\sigma_{3-9} = 0.25 \pm 0.02$, and $\sigma_{10-13} = 0.60 \pm 0.10$.  For both the LMC and WLM there appears to be an increase in the metallicity spread for older ages, however as pointed out by \cite{Cole05} this can be due to systematic age effects blurring the age-metallicity relation.  In the case of the LMC the rapidly rising [Fe/H] enrichment in the first few Gyrs also adds to the high intrinsic spread in [Fe/H] at large ages.  However the AMR is relatively flat in early times for WLM, so the increased [Fe/H] dispersion at large ages could be linked to a radial mixing process.  Both the LMC and WLM also show an increase in velocity dispersion with age, as well as increase in metallicity dispersion with age, however modelling of such joint chemodynamic signatures is out of the scope of this paper.

\section{Discussion}
In the following section we asses the likelihood that secular or environmental factors could produce the particular similarities and differences observed in the metallicities, structure, and dynamics of the sample of galaxies.  

\subsection{Global Metallicity Properties in the Sample}
Despite WLM's isolation, Figures \ref{fig:difcomp} and \ref{fig:siglums} suggest that its bulk chemical properties are similar to the more luminous of the tidally disturbed dSphs and the Magellanic Clouds. Approximately the same mean metallicity is found in WLM, Fornax, Leo I and the SMC, while their tidal indices\footnote{\cite{Karachentsev05} define $\Theta \equiv {\rm max [log}(M_{k}/D_{ik}^{3}){\rm ] + C}$ to be the amount a galaxy is acted on by its largest tidal disturber. As noted in that work, $\Theta = 0$ corresponds to an object with a Keplerian orbital period about MW/M31 equal to $1/H$.  For reference, Sgr dSph has $\Theta = 5.6$.} range from $\Theta = 0.2$ to $\Theta = 3.5$\footnote{If Fornax became a satellite of the MW relatively recently or, as is suggested by proper motion measurements, has a circular orbit about the MW, similarities in chemical evolution may be expected despite the current present day environmental differences.} \citep{Karachentsev05}.  Stars are found as metal poor as [Fe/H]$ \simeq -2.9$ in WLM, the same lower limit as Fornax, suggesting that dSphs and dIrrs of the same luminosity have stars of low metallicity present - consistent with the survey of \cite{Weisz11}.  Together with the similar AMR shown for these galaxies in Figure \ref{fig:wlmamr}, the comparative metallicity properties suggest that the chemical enrichment proceeds largely independent of environment and is primarily dictated by the mass of the galaxy - in agreement with the simulations of \cite{Sawala11}.  This is also consistent with the recent observations of the isolated dwarf galaxy VV124 by \cite{Kirby12}, which was found to lie on the Local Group MMR despite its extreme isolation. 

Similarities in the AMR and chemical enrichment timescales of the SMC, WLM and Fornax may also be indicative of a common \emph{initial} mass - with subsequent differences in the present day masses plausibly attributed to tidal stripping by the Milky Way.  Using estimates for the half light masses from \cite{JWolf10} and Paper II, in conjunction with the tidal evolutionary tracks of \cite{Penarrubia08}, it can be shown that if the SMC were to have lost $\sim 40\%$ and $\sim 98\%$ of its total mass, it would have a similar half light mass as WLM and Fornax respectively.  Similarly were WLM to undergo stripping of $\sim 90\%$ of its total mass, it would have a comparable present day mass to Fornax.  It is certainly plausible then that the similar enrichment history of those three galaxies is linked to similar infall masses.  Adding weight to this argument is the presence of GCs in both WLM and Fornax, as well as planetary nebulae \citep{Magrini05,Kniazev07}, which may imply common masses and SFH (cf., \citealt{Saviane09}).

The measured global metallicity properties of WLM are in good agreement with the expectations for its luminosity, regardless of environment.  WLM shows no discernible offset or trend with respect to the luminosity metallicity (LZR) relation of \cite{Kirby08}.  The offset between dIrrs and dSphs reported by \citep{Grebel03,Woo08} does not appear as strongly, if at all, when plotting the updated metallicities for the galaxies we consider.  Such systematic offsets are difficult to confirm and interpret in cases where biased metallicity indicators or small spatial coverage skew the average [Fe/H] values.  WLM's agreement with the dSph LZR is consistent with studies looking at large samples of more distant galaxies from the SDSS survey - where only small deviations from the MZR are seen for galaxies of different environment or morphological class \citep{Ellison08,Ellison09}.  Environmental processing such as ram pressure stripping may not necessarily produce offsets in any one particular direction from the mass metallicity relation, however there is still likely a complex interplay between gas stripping, triggered star formation and dilution of the ISM during infall of a dwarf galaxy to a larger spiral.  This may add to the observed scatter in the LZR \citep{Skillman96,Boselli08,Koleva11}.  In addition varying levels of angular momentum in the dwarfs may modulate SF efficiency \citep{Schroyen11}, with the low angular momentum galaxies having more centrally concentrated gas which may be more efficient at cooling into the molecular phase \citep{Kuhlen11}.

The spread in metallicity, $\sigma_{\rm{[Fe/H]}}$, for the dwarf galaxies studied here is nearly constant over 4 orders of magnitude in luminosity, with a mean of $\langle\sigma_{\rm{[Fe/H]}}\rangle = 0.38$.  The anticorrelation with luminosity found for the ultra faint dwarfs seems to saturate at this value for galaxies with luminosity $L \gtrsim 5\times10^{5}$ L$_{\odot}$ in Figure \ref{fig:siglums}.  If the spread in [Fe/H] reflects the stochastic nature of enrichment events, then perhaps higher luminosity systems (where SF proceeds over several Gyr and many supernovae occur) end up with a similar value of $\sigma_{\rm{[Fe/H]}}$ which reflects the mixing efficiency in the ISM \citep{Argast00}. Lower luminosity systems which have truncated SFHs and therefore fewer enrichment events will unevenly distribute metals through the galaxy in the short time they are forming stars leading to the higher dispersion [Fe/H].

\subsection{Spatial Abundance Signatures of the Sample}
Past studies \citep{Winnick03,Koch07,Gullieuszik09,Kirby11a} have found both flat and statistically significant negative radial [Fe/H] profiles for dSphs.  However, in studies using linear CaT calibrations unbiased gradient estimates have proved difficult to compute.  Even then small spatial coverage may miss such a gradient. As pointed out by \cite{Kirby11a}, their sample of stars in Fornax showed a significantly different gradient from the DART sample simply due to the small region of the galaxy sampled - which is reinforced when looking at the inner regions of our Figure \ref{fig:rellover}. Due to the strong RGB metallicity gradients shown by the dSphs, it is necessary to sample out to large radii in order to detect the low metallicity stars and build up a representative MDF.  

In this work we have selected literature data which show the least bias in terms of spatial coverage and [Fe/H] calibration to allow for a consistent differential comparison of the chemical trends with radius as traced by RGB stars. With the unbiased metallicity and spatial coverage in the sample we consider, there appears to be a slight dichotomy in the abundance gradients of dIrrs and dSphs - with the gas rich rotating dIrrs showing significantly shallower radial gradients.  While this could be due to differences in their total masses, further evidence supporting angular momentum as a driver of gradient strength comes when looking at the Local Group dEs: NGC 205, NGC 185, and NGC147.  The dispersion dominated system NGC 205 ($V/\sigma \sim 0.3$; \citealt{Geha06}) was found by \cite{Koleva09} to have a metallicity gradient comparable to the dSphs in our sampler ($\rm{d[Fe/H]/d{r_{c}}} \gtrsim -0.14$ dex r$_{\rm{c}}^{-1}$).  By contrast NGC 147 and NGC 185 ($V/\sigma \geq 0.91, V/\sigma \geq 0.65$; \citealt{Geha10}) were found to have gradients as flat or flatter than the dIrrs we consider here ($\rm{d[Fe/H]/d{r_{c}}} \simeq -0.02$ dex r$_{\rm{c}}^{-1}$; \citealt{Geha10}).

To quantify this, in Figure \ref{fig:angvfeh} we plot the slope of the radial metallicity gradient versus $V/\sigma$.   $V/\sigma$ is not a singular quantity, but changes with radius within a galaxy - therefore we show the range of $V/\sigma$ spanned by particular galaxies in this plot.  The measured values for the dSphs (and the LMC) must be taken as coarse estimates, and should be interpreted with care.  This is due to several factors: in the case of nearby dSphs with large angular extents it can be difficult to ascertain whether the observed velocity gradients are intrinsic rotation or perspective induced rotation, the uncertain inclination and axis of rotation for the dSphs may lead to an underestimate of any true rotational velocity \citep{Lokas10}, and any tidal distortions to the dSphs make interpretation of their dynamical state difficult. Nevertheless there is evidence for the lowest angular momentum systems to show steeper metallicity gradients in this literature sample.

It should be kept in mind that when comparing radial gradients (especially in systems with ongoing star formation), ideally a tracer population of stars of the same age should be used.  Indeed the change in abundance gradient as a function of population age offers a strong tool to study the chemical evolution \citep{Cioni09,Vlajic09}.  For example the LMC shows a decrease in $V/\sigma$ with age, and the older stars are also more spatially extended. Therefore one would want to estimate the kinematics, chemistry and scale length of a single age stellar population in order to accurately test chemodynamic signatures - however this is extremely difficult to do with current data of Local Group dwarfs.  The LMC may be a very complex case which should be analyzed with caution, however the similar metallicity gradients, dynamical state, mass and size of WLM, SMC, NGC 147, and NGC 185 certainly offer a robust comparison sample to the dispersion dominated dSphs.  While the dSphs may show on average older RGB stars, we note that WLM and Fornax have nearly identical age metallicity relations (Fig. \ref{fig:wlmamr}), yet show the most disparate radial abundance gradients.  

\subsubsection{The Role of Environment} 
\begin{figure}
\begin{center}
\ifpdf
\includegraphics[width=0.9\textwidth]{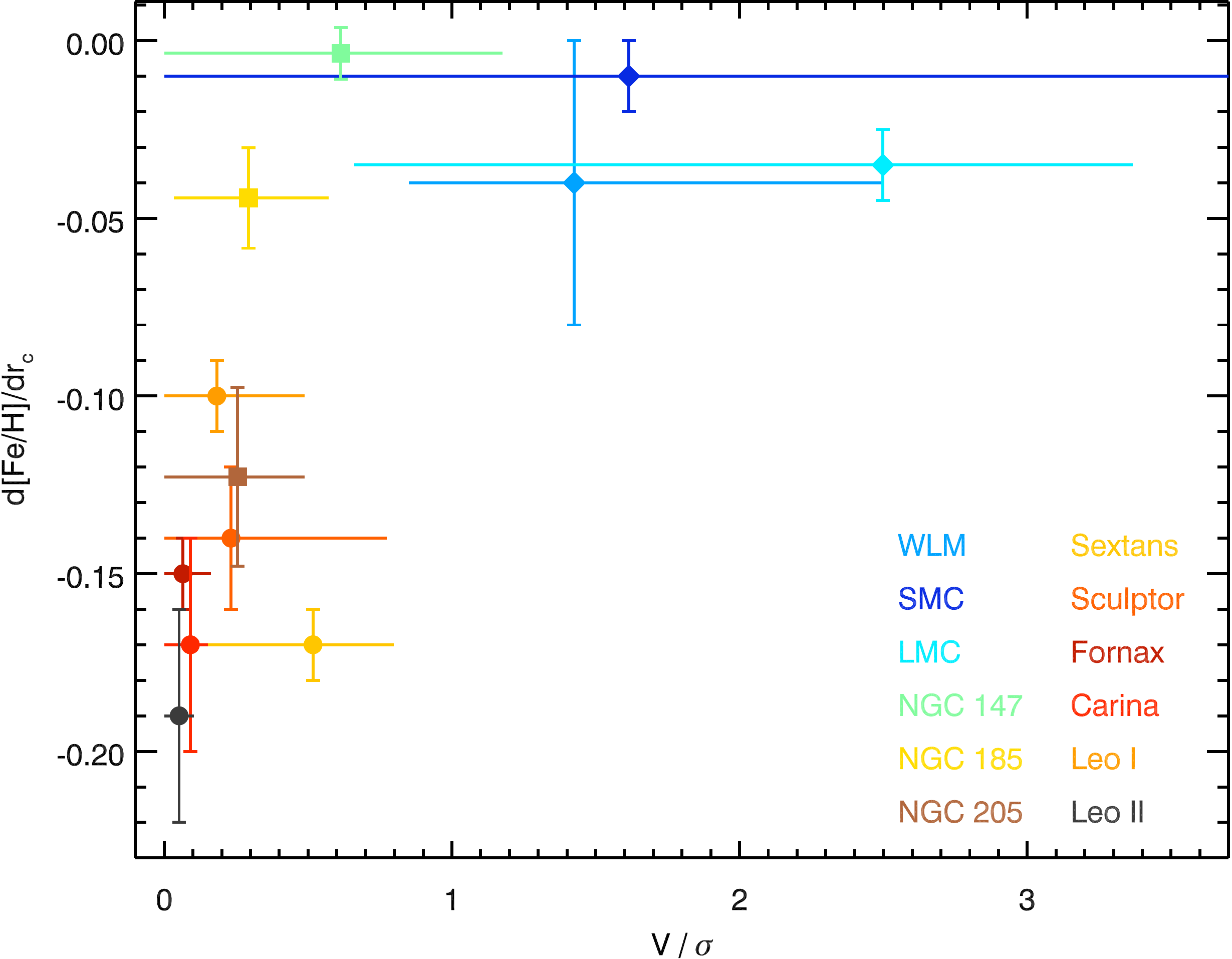}
\else
\includegraphics[angle=90,width=0.9\textwidth]{angvfeh2.eps}
\fi
\caption{The radial metallicity gradients plotted against the ratio of rotation to pressure support.  The $V/\sigma$  lines in the x-direction are not the errors, but show the range spanned by the galaxy over the radii where kinematic and metallicity information exist, and should be considered coarse estimates.  In the dSphs it should be stressed that the observed rotation may not be intrinsic rotation, or may be overestimated due to uncertain kinematic axes or tidal features, and should be considered coarse estimates only (see text).  Data for the velocity information comes from \cite{Mateo08} (Leo I), \cite{Koch07} (Leo II), \cite{Walker09} (Fornax, Carina), \cite{Battaglia08} (Sculptor), \cite{Battaglia11} (Sextans), \cite{Geha10} (NGC 147, NGC 185), \cite{Geha06} (NGC 205), \cite{HZ06} (SMC), and \cite{vdM02} (LMC). }
\label{fig:angvfeh}
\end{center}
\end{figure}

The RGB gradients in Figure \ref{fig:rellover} and \ref{fig:angvfeh} for the gas rich, rotating dwarf galaxies are shallower than the low angular momentum gas poor dSphs.  However there are not as large differences between the metallicity gradient of the isolated WLM galaxy and the tidally disturbed Magellanic Clouds.  If the metallicity gradient is modulated by environmental factors such as ram pressure or tidal stripping, this would be consistent with recent observational and theoretical work suggesting that the Magellanic Clouds are on their first passage into the MW virial radius \citep{Besla07,BK11}.  Despite their close present day distance to the MW, their gas content and spatially extended SF may have been undisturbed, allowing chemical enrichment to proceed over the full body of the LMC and SMC for 10+ Gyr.  In this environment driven scenario, the steeper metallicity gradients of dSphs would be due to their early infall times - with their gas content and chemical enrichment quenched as they were accreted by the MW.  This could perhaps lead to a radially shrinking SF region (and steeper metallicity gradient) within those dwarfs (i.e, \citealt{Mayer07, Mayer11}).

In addition to ram pressure mediating metallicity gradients, it is not clear what factor tidal stripping of the stellar populations will play in preserving or erasing a gradient.  As shown by \cite{Sales10} the kinematic and metallicity gradients in dSphs near the MW may be subject to tidally induced modifications depending on their initial strength and the orbital properties of the dSph.  In most cases however, the metallicity gradient would remain, as the outer most metal poor stars would be unbound first.  Therefore in situ formation of gradients for dSphs could still be preserved (or made milder) in the presence of environmental processing from the MW (see also \citealt{Koleva11}).  The trends in Figure \ref{fig:angvfeh} may present a test for merger or tidally induced transformations of dIrrs into dSphs (e.g, \citealt{Mayer01a,Klimentowski09,Kazantzidis11,Kazantzidis11b}) - as in such scenarios it could be difficult to produce objects with steep metallicity gradients from rotating progenitors which had flat radial metallicity profiles.  Ram pressure or additional baryonic effects may allow for more flexibility \citep{Mayer07,Kravtsov10} on this issue (and on the evolution of the dwarfs' M/L ratios).  Further modelling which jointly treats the chemical and kinematic evolution of dwarf galaxies in the MW potential could comment directly.

\subsubsection{Internal Processes} 
Simulations by \cite{Schroyen11} showed that angular momentum plays a strong role in determining the radial metallicity profile of a dwarf galaxy.  In their simulations rotation produced a centrifugal barrier which in turn prevented gas from settling in the centre of galaxies.  This led to SF that occurred over the full extent of the dwarf at lower levels.  This scenario naturally produced smoother radial metallicity profiles, and extended star formation histories.  These results were in qualitative agreement with the observations of dwarf ellipticals by \cite{Koleva09}, but a larger sample by \cite{Koleva11} from the literature has found weaker correlations with host galaxy properties.  One simulation run from \cite{Schroyen11} shows a metallicity gradient of $\rm{d[Fe/H]/d{r_{e}}} \sim -0.12$ dex r$_{\rm{e}}^{-1}$ for the non-rotating dwarf, with the rotating dwarf ($V/\sigma \sim 1.8$) having a gradient of $\rm{d[Fe/H]/d{r_{e}}} \sim -0.03$ dex r$_{\rm{e}}^{-1}$.  This is in excellent agreement with the average radial gradients we find for the dSphs ($-0.13\pm0.01$) and dIrrs ($-0.03\pm0.01$).  Together with the dEs discussed above these observations would support a correlation between angular momentum and radial metallicity gradient strength in Local Group dwarf galaxies.

Another secular process for flattening metallicity gradients demonstrated in simulations \citep{SellwoodBinney02,Roskar08,Stinson09,Loebman11} is radial migration of stars.  The migrations of the stellar populations are typically produced by global disk instabilities or transient spiral structures.  These are thought to be common in MW sized galaxies, however it is unclear how ubiquitous such dynamical instabilities are in low mass, thickened dwarf galaxies \citep{Sotnikova03,SJ10,Mayer11}.  Redistribution of locally enriched material within a galaxy is also possible due to the SF driven fountain mechanism as shown by \cite{DeYoung94}, but again this may not be dominant in low mass, thick dwarfs \citep{Schroyen11}.  If the dIrrs were high enough mass for one or both of these processes to operate, the dichotomy in metallicity gradients may then be due to total dynamical mass - however the similar masses of NGC 205, 185, and 147 would argue against this.

In summary there could be several mechanisms at work to produce the observed radial [Fe/H] gradients.  More data are needed to reveal whether environment, total mass, or angular momentum is the more fundamental parameter modulating metallicity gradients in dwarfs - which may be difficult as a correlation between $V/\sigma$ and environment may also be present. Efforts to incorporate metallicity profiles into dynamical simulations of DG evolution may also increase constraints on the contribution of environment.  In all cases it is imperative to ascertain the dynamical and chemical profiles of the dwarfs out to large radii - as the value of $V/\sigma$ in particular will change dramatically from the inner to outer regions.

\section{Summary}
In this paper we have presented [Fe/H] and age estimates for 126 RGB stars in the WLM dIrr galaxy.  These complement the kinematic and structural study of WLM presented in our earlier papers, and represent some of the first spectroscopic abundances and velocities of individual evolved stars in a truly isolated Local Group dwarf galaxy.  
For WLM and the literature data we have computed the metallicity distribution functions and characterized the bulk chemical abundance properties, as well as calculated the radial metallicity gradients, and age-metallicity relation for each galaxy as traced by the RGB stars.

The key points from our study are:
\begin{itemize}
\item The global metal abundances (mean, median) and MDF for the isolated dwarf galaxy WLM are similar to the SMC and the more luminous dSph Fornax  - in expectation with the luminosity-metallicity relation for the Local Group.
\item WLM shows stars at as low metallicity as the dSph Fornax, however the dSphs on average show more extended metal poor tails, and are better fit by simple leaky box chemical evolution models than the dIrrs we consider.
\item The intrinsic spread in a galaxy's metallicity is constant over 4 orders of magnitude in luminosity, suggesting that the dispersion in [Fe/H] may saturate for dwarf galaxies with luminosities L$\gtrsim 5 \times 10^{5}$ L$_{\odot}$.
\item The dispersion in metallicity increases with age in WLM as is found in the LMC - with both galaxies also showing an increase in velocity dispersion in age (Paper II).
\item WLM, along with the SMC and LMC show radial [Fe/H] profiles that are statistically flatter than the dSphs. This along with the flat gradients for the rotating dEs NGC 147 and NGC 185 supports a dichotomy in radial metallicty gradients which correlates with angular momentum for Local Group dwarfs.
\end{itemize}

The strength of this study lies in the use of spatially extended samples of stars which have unbiased [Fe/H] indicators.  Further measurements of stellar spectra at large radii may reveal as yet undiscovered velocity and metallicity gradients in dSphs or dIrrs which could help confirm the chemodynamic correlations suggested here.  If borne out by further observations, such correlations between angular momentum and radial metallicity gradients could offer useful constraints for models of environmentally driven transformations of dIrrs into dSphs. Specific simulations tracking the angular momentum and radial chemical enrichment of an infalling dSph progenitor would be useful to improve the current understanding of how the classes of dwarf galaxies are connected.  This coupled with additional spatially complete surveys of the chemistry in isolated dIrrs as well as dSphs will undoubtedly shed light on the magnitude of internal and environmental effects that shape the morphologies and chemistries of the Local Group dwarf populations.

\acknowledgments
We are indebted to the referee, Dr. Ivo Saviane, for thorough readings and many useful comments and suggestions which greatly improved this manuscript.  RL acknowledges support from NSERC Discovery Grants to Don VandenBerg and KV, and acknowledges financial support to the DAGAL network from the People Programme (Marie Curie Actions) of the European Union’s Seventh Framework Programme FP7/2007- 2013/ under REA grant agreement number PITN-GA-2011-289313.  The authors acknowledge the International Space Science Institute (ISSI) at Bern for their funding of the team ``Defining the full life-cycle of dwarf galaxy evolution: the Local Universe as a template'', as well as useful discussions with associated team members.  The research leading to these results has received funding from the European Union Seventh Framework Programme (FP7/2007-2013) under grant agreement number PIEF-GA-2010-274151.  AB acknowledges support from the Sherman Fairchild Foundation and The Grainger Foundation.  RI gratefully acknowledges support from the Agence Nationale de la Recherche though the grant POMMME (ANR 09-BLAN-0228).  ES and ET gratefully acknowledge Netherlands Foundation for Scientific Research (NWO) for financial support.  The authors would like to thank M. Shetrone, A. Ferguson, R. L\"asker, S. Ellison, K. Bekki, J. Pe\~{n}arrubia and G. van de Ven for useful discussions, and R. Carrera, M. Parisi, L. Pomp\'{e}ia, E. Kirby, M. Geha, and A. Koch for making public their data.  FORS2 observations were collected at the ESO, proposal 072.B-0497.  Additional data presented herein were obtained at the W.M. Keck Observatory, which is operated as a scientific partnership among the California Institute of Technology, the University of California and the National Aeronautics and Space Administration. The Observatory was made possible by the generous financial support of the W.M. Keck Foundation.  The authors wish to recognize and acknowledge the very significant cultural role and reverence that the summit of Mauna Kea has always had within the indigenous Hawaiian community.  We are most fortunate to have the opportunity to conduct observations from this mountain.
\textit{Facilities:} \facility{VLT:Yepun(FORS2)}, \facility{Keck:II(DEIMOS)}

\bibliography{wlmrefst}

\clearpage

\clearpage

\clearpage

\clearpage

%

\clearpage

\clearpage

\clearpage

\clearpage

\clearpage
\begin{deluxetable}{cccl}
\tablecolumns{4}
\tablewidth{0pc}
\tablecaption{Local Group Dwarf Galaxy Sample}
\tablehead{
\colhead{Galaxy} & \colhead{N$_{\rm{stars}}$} & \colhead{$r_{max}/r_{t}$}\tablenotemark{a} & \colhead{Reference}
}
\startdata
WLM & 180 & 0.79 & \cite{Leaman09,Leaman12}\\
LMC & 373, 59, 383 & 0.78 & \cite{Cole05,Pompeia08,Carrera08b}\\
SMC & 349, 364 & 1.14 & \cite{Carrera08, Parisi10}\\
Fornax & 870 & 1.20 & \cite{Battaglia06}\tablenotemark{b}\\
Sculptor & 629 & 1.34 & \cite{Tolstoy04}\tablenotemark{b}\\
Sextans & 180 & 0.75 & \cite{Battaglia11}\tablenotemark{b}\\
Leo I & 825 & 1.18 & \cite{Kirby10}\\
Leo II & 256 & 1.10 & \cite{Kirby10}\\
Carina & 327 & 1.06 & \cite{Koch06}\tablenotemark{b}\\
\enddata
\tablenotetext{a}{Column shows what fraction of the tidal radius the outer most star in the spectroscopic sample extends to.}
\tablenotetext{b}{``DART sample'' - original data from these papers updated with additional observations and the [Fe/H] calibration from \cite{Starkenburg10}.}
\end{deluxetable}

\begin{deluxetable}{cccccc}
\tablecolumns{6}
\tablewidth{0pc}
\tablecaption{MDF Properties}
\tablehead{
\colhead{Galaxy} & \colhead{10$^{\rm{th}}$ \%} & \colhead{50$^{\rm{th}}$ \%} & \colhead{90$^{\rm{th}}$ \%} & \colhead{$p$  [$Z_{\odot}$]} & \colhead{[Fe/H]$_{0}$}
}
\startdata
LMC & $-1.06$ & $-0.45$ & $-0.18$ & $0.430$ & $-\infty$\\
    &       &       &       & $0.363$ & $-1.30$\\[1mm]
\hline\\[-1.8mm]
SMC & $-1.53$ & $-1.05$ & $-0.64$ & $0.100$ & $-\infty$\\
    &       &       &       & $0.085$ & $-1.91$\\[1mm]
\hline\\[-1.8mm]
WLM & $-1.74$ & $-1.24$ & $-0.75$ & $0.070$ & $-\infty$\\
    &       &		&		& $0.064$ & $-2.34$\\[1mm]
\hline\\[-1.8mm]
Fornax & $-2.04$ & $-1.17$ & $-0.74$ & $0.093$ & $-\infty$\\
	   &	   &	   &       & $0.090$ & $-2.66$\\[1mm]
\hline\\[-1.8mm]
Leo I & $-1.84$ & $-1.42$ & $-1.11$ & $0.090$ & $-\infty$\\
	  & 	  &       &       & $0.037$ & $-2.18$\\[1mm]
\hline\\[-1.8mm]
Sculptor & $-2.45$ & $-1.96$ & $-1.41$ & $0.014$ & $-\infty$\\
		 &		 &		 &		 & $0.013$ & $-3.14$\\[1mm]
\hline\\[-1.8mm]
Leo II & $-2.29$ & $-1.59$ & $-1.28$ & $0.036$ & $-\infty$\\
	   &	   &	   &	   & $0.033$ & $-2.68$\\[1mm]
\hline\\[-1.8mm]
Sextans & $-2.89$ & $-2.26$ & $-1.66$ & $0.007$ & $-\infty$\\
	    & 		& 		& 		& $0.007$ & $< -5.0$\\
\hline\\[-1.8mm]
Carina & $-2.47$ & $-1.87$ & $-1.51$ & $0.019$ & $-\infty$\\
	   & 	   &	   &       & $0.017$ & $-3.02$\\[-1.8mm]
\enddata
\tablecomments{Effective yields in the first row for each galaxy represent the best fitting value from a Leaky box model, second row the effective yield in the pre-enriched model and initial [Fe/H].}
\end{deluxetable}

\clearpage

\end{document}